\documentclass[prb,twocolumn]{revtex4}
\usepackage[utf8]{inputenc}
\usepackage{bm}
\usepackage{graphicx}
\usepackage{amsmath}
\usepackage{color}
\usepackage{upgreek}
\usepackage{hyperref}
\usepackage[normalem]{ulem} 
\usepackage{booktabs}
\usepackage{tabularx}
\usepackage{textcomp}

\newcolumntype{C}{>{$}c<{$}}
\AtBeginDocument{
\heavyrulewidth=.08em
\lightrulewidth=.05em
\cmidrulewidth=.03em
\belowrulesep=.65ex
\belowbottomsep=0pt
\aboverulesep=.4ex
\abovetopsep=0pt
\cmidrulesep=\doublerulesep
\cmidrulekern=.5em
\defaultaddspace=.5em
}
\begin{document}

\title{Cyclotron resonance induced photogalvanic effect in  surface states of 200 nm thick strained HgTe films}

\author{S.\,Candussio,$^1$ G.V.\,Budkin,$^2$ M.\,Otteneder,$^1$  D.A.\,Kozlov,$^{3,4}$ 	 I.A. Dmitriev,$^{1,2}$ V.V.\,Bel'kov,$^2$ Z.D.\,Kvon,$^{3,4}$   N.N.\,Mikhailov,$^3$ S.A.\,Dvoretsky,$^3$ and S.D.\,Ganichev$^1$}

\affiliation{$^1$Terahertz Center, University of Regensburg, 93040 Regensburg, Germany}

\affiliation{$^2$Ioffe Institute, 194021 St.\,Petersburg, Russia}

\affiliation{$^3$A.V. Rzhanov Institute of Semiconductor Physics, Novosibirsk 630090, Russia,}

\affiliation{$^4$Novosibirsk State University, Novosibirsk 630090, Russia }



\begin{abstract}
We report on the study  of magneto-photogalvanic and magnetotransport phenomena in 200\,nm partially strained HgTe films. This thickness is slightly larger than the estimated critical thickness of lattice relaxation leaving the film partially relaxed with the value of the energy gap close to zero. We show that illumination of HgTe films with  monochromatic terahertz laser radiation results in a giant resonant photocurrent caused by the cyclotron resonance in the surface states. The resonant photocurrent is also detected in the reference fully strained 80\,nm HgTe films previously shown to be fully gapped 3D topological insulators. We show that the resonance positions in both types of films almost coincide demonstrating the existence of topologically protected surface states in thick HgTe films. The conclusion is supported by magnetotransport experiments.
\end{abstract}
\maketitle

\section{Introduction}

Mercury cadmium telluride structures recently attracted growing attention as a material for fabrication of the highest quality two- and three-dimensional topological insulators (TIs).\cite{ColloquiumTI} Bulk mercury telluride is known to be gapless, therefore, in order to obtain a topological insulating state one uses either size quantization in quantum well structures (for 2D TIs\cite{Bernevig2006}) or tensile strain to open a gap (3D TIs\cite{Brune2011}). The strain naturally appears when HgTe films are grown on CdTe substrates having 0.3\% lattice mismatch. It was shown that  70-80\,nm thick HgTe films adopt the CdTe lattice constant, which results in an energy gap of around 15\,meV.\cite{Brune2011, Kozlov2014} A characteristic feature of such films is the existence of topologically protected surface states. Their presence was first demonstrated by observation of the spin quantum Hall effect.\cite{Brune2011} The properties of surface states have been comprehensively studied using  magnetotransport\cite{Brune2011,Kozlov2014,Wiedenmann2016, Wiedenmann2017, Maier2017, Ziegler2018} and capacitance spectroscopy,\cite{Kozlov2016} as well as optical and photogalvanic spectroscopy.\cite{Shuvaev2012, Shuvaev2013a, Dantscher2015, Wu2016} Though study of thicker HgTe films is also of indubitable interest, the requirement of strain produced by the lattice mismatch implies the existence of a critical thickness above which the gap should close since the strain relaxes due to formation of dislocations. According to previous works, the critical film thickness of HgTe grown on CdTe is expected to lie between $100$ and $200$\,nm.\cite{Brune2011, Kozlov2014} The questions are whether in such relaxed films the topological surface states are still present and how their properties are affected by the presence of bulk carriers. Very recent magnetotransport measurements\cite{Savchenko2018} combined with capacitance spectroscopy provide a first clear indication of the presence of surface states in 200\,nm partially relaxed HgTe films with almost zero bulk energy gap.

Here we report on the study of magneto-photogalvanic phenomena supported by temperature-dependent magnetotransport measurements in 200\,nm HgTe films. Our results demonstrate that, while the thickness of these films is slightly larger than the estimated critical thickness of lattice relaxation, the surface states with characteristics similar to the  fully strained 80\,nm films can be clearly detected. In particular, both for 200 and 80\,nm films we observe the cyclotron resonance (CR) induced photogalvanic effect exhibiting two resonant features characterized by slightly different cyclotron masses and corresponding to the top and bottom surface states. The positions of resonances in the photovoltage coincide with the CR dips in simultaneously measured magnetotransmission. The experimental results and analysis of the THz photocurrent and magnetotransmission are complemented by magnetotransport measurements.

\section{Samples, magnetotransport data and experimental setup}

\begin{figure}[t]
\begin{center}
\includegraphics[width=\linewidth]{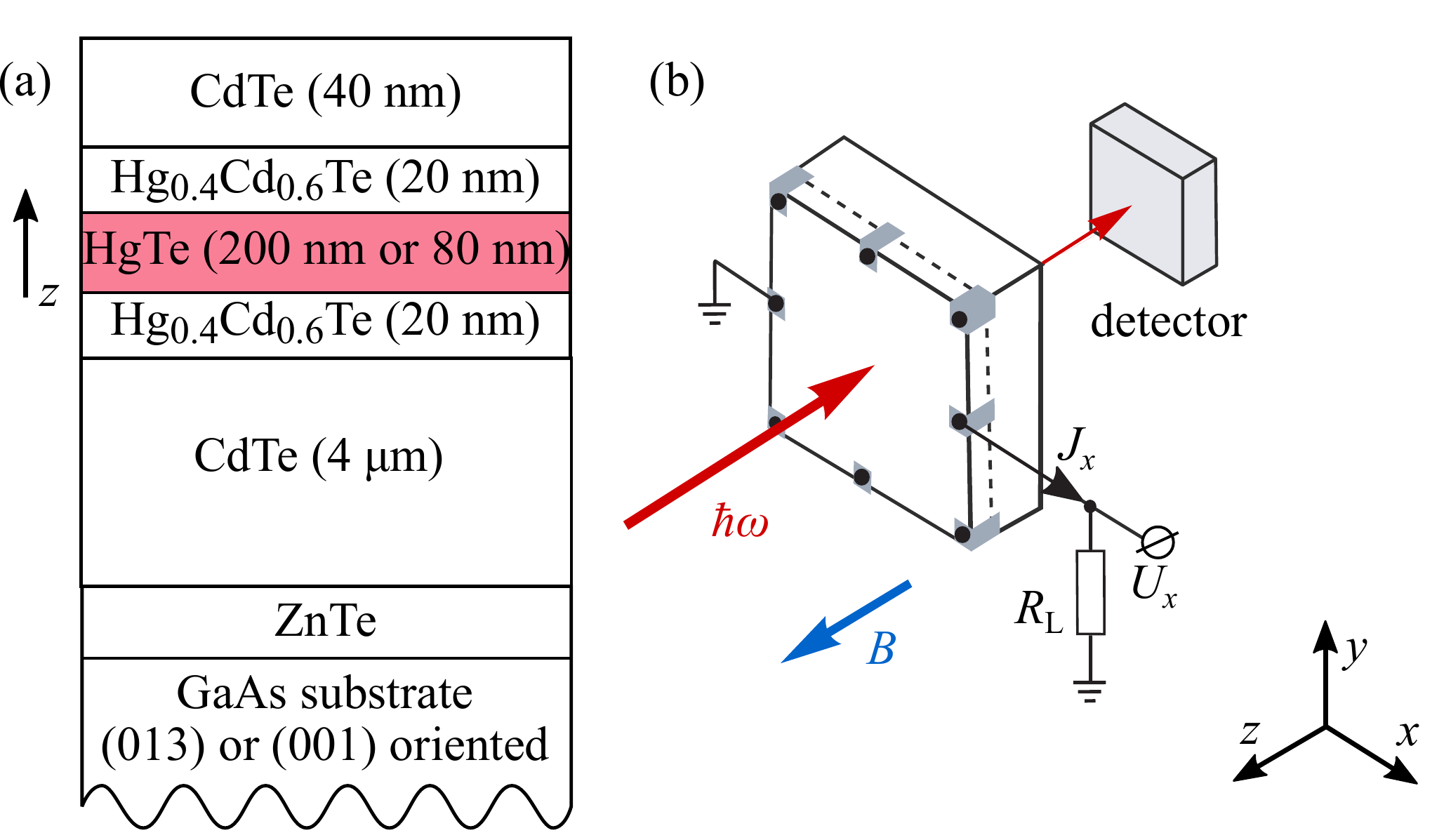}
\caption{Panel (a) Cross-section of the investigated structures. The $200$ and $80$\,nm thick HgTe films were grown on (013)- and (001)-oriented GaAs substrates. (b) Experimental setup of the photocurrent and radiation transmission measurements.}
\vspace{-1cc}
\label{fig1}
\end{center}
\end{figure}

Our experiments have been carried out on high-mobility HgTe films with a thickness of $200$ and $80$\,nm grown via molecular beam epitaxy. Here, the HgTe film is embedded between two layers of Cd$_{0.6}$Hg$_{0.4}$Te which work as cap and buffer layer. The cross section of the structure is sketched in Fig.~\ref{fig1}(a). The structures with 200\,nm films are fabricated on (013)-oriented GaAs substrates and the reference 80\,nm films on either (013)- or (001)-oriented substrates.  Note that 200\,nm films processed from the same wafer were also used for magnetotransport measurements in Ref.~[\onlinecite{Savchenko2018}]. The films were grown on a 4\,$\upmu$m thick, fully relaxed CdTe layer deposited on a thin ZnTe layer, see Fig. \ref{fig1}(a). Square shaped $5 \times 5$\,mm$^2$  samples have been fabricated, with one of the edges oriented along $x \parallel [100]$ being a cleaved edge face. This geometry allows us to measure radiation transmission through the sample and THz radiation induced photocurrents simultaneously. For the photocurrent and magnetotransport measurements, four ohmic contacts in the corners and four in the middle of the sample edges have been prepared, see Fig. \ref{fig1}(b). To vary the Fermi level position some of the 200~nm structures were equipped with semitransparent NiCr (14~nm thick) gates on top of 100~nm Al$_2$O$_3$ films grown by atomic layer deposition.

\begin{figure}[t]
\begin{center}
\includegraphics[width=\linewidth]{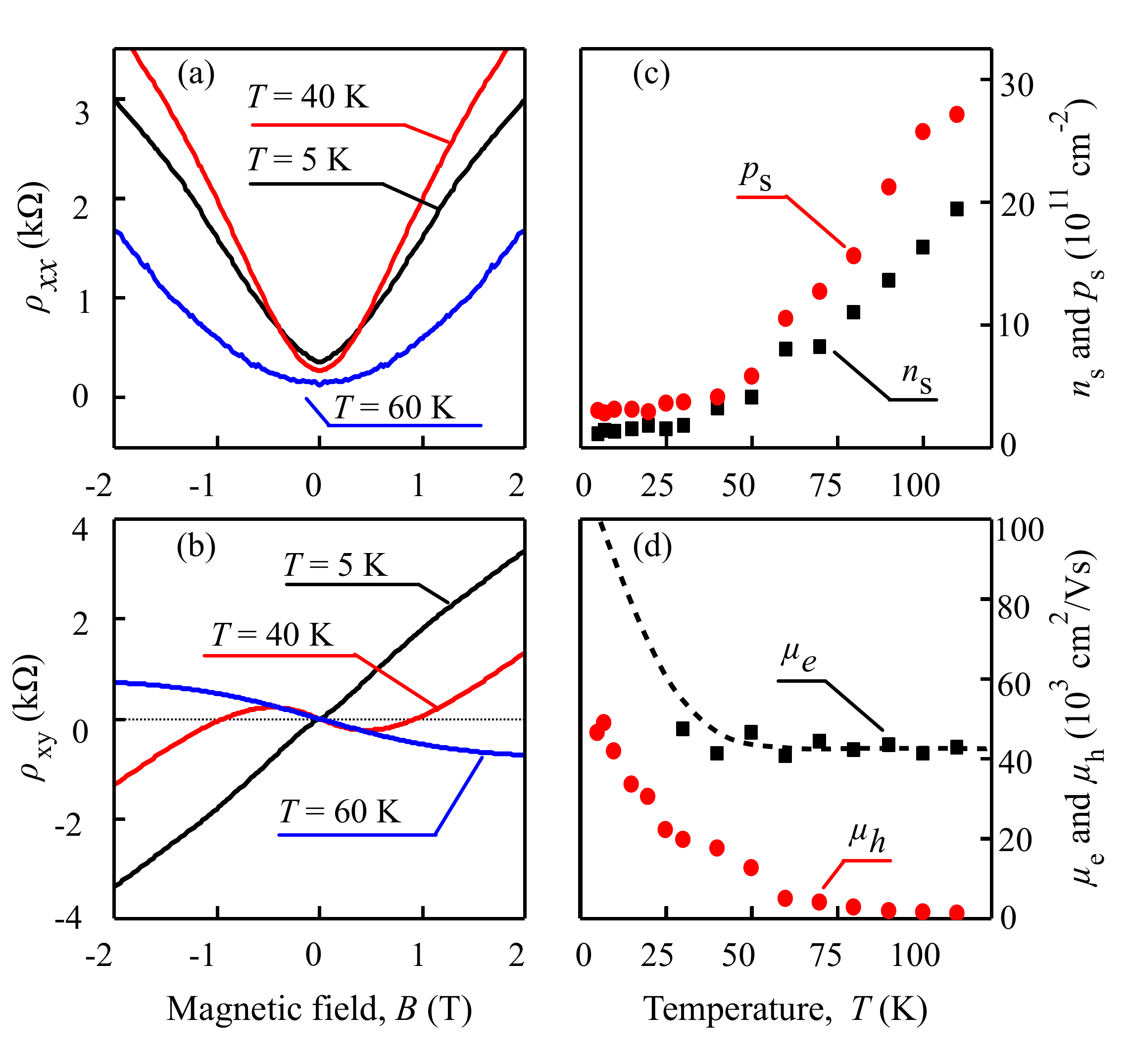}
\caption{Magnetotransport data for the $200$\,nm HgTe film. Panels (a) and (b) show the magnetic field dependencies of the longitudinal sheet resistance $\rho_{xx}(B)$ and the Hall resistance $\rho_{xy}(B)$ measured on a square-shaped van der Pauw sample. Panels (c) and (d) display the temperature dependencies of electron and hole densities $n_s$ and $p_s$, respectively, as well as the corresponding mobilities $\mu_e$ and $\mu_h$. The dependencies were extracted via fitting of $\rho_{xx}(B)$ and $\rho_{xy}(B)$ traces using classical Drude theory. We do not present fitted values for the electron mobility at low $T$, where the procedure is not reliable because of low electron density. An expected mobility dependence is shown by the dashed line.}
\label{fig2}
\vspace{-1cc}
\end{center}
\end{figure}

The magnetotransport measurements were performed in magnetic field $B$ up to 7\,T oriented normal to the HgTe film using standard low-frequency lock-in technique with a driving current of 1\,$\upmu$A. The obtained longitudinal ($\rho_{xx}$) and Hall ($\rho_{xy}$) sheet resistances of the $200$\,nm HgTe film, shown in Figs.~\ref{fig2}(a) and \ref{fig2}(b) for selected temperatures in the range from $4.2$ to $110$\,K, display behavior similar to the previously investigated $80$\,nm films,\cite{Kozlov2014,Dantscher2015,Savchenko2018} with the nonlinear $N$-type Hall resistance indicating the coexistence of electrons and holes. The change of sign of the Hall resistance with temperature at high $B$ indicates thermal activation of carriers. The hole-dominated transport at low temperature is in accordance with previous works revealing that in ungated HgTe films the Fermi level is effectively pinned to the valence band.\cite{Kozlov2014,Dantscher2015} The electron contribution to the conductivity at low temperatures is attributed to surface states, and at high temperature is enhanced by an increased fraction of bulk electrons. Using classical two-carrier Drude model,\cite{Kozlov2014, Kvon2008} from $\rho_{xx}(B)$ and $\rho_{xy}(B)$ we extracted densities and mobilities of electrons and holes in the system. The obtained temperature dependencies are shown in panels (c) and (d) of Fig.~\ref{fig2}.

Photocurrent and magnetotransmission measurements have been performed applying radiation of a continuous wave ($cw$) optically pumped molecular gas terahertz (THz) laser.\cite{Ganichev2009,Olbrich2011,Olbrich2014,Ganichev1982} The laser lines with frequencies of $f=2.54$, $1.62$, and $0.69$\,THz (wavelengths $\lambda=118$, $184$, and $432\,\upmu$m) were obtained using methanol, diflouromethane, and formic acid as an active media, respectively. The incident power, $P$, depends on the particular laser line and ranges from $5$\,mW for $f=0.69$\,THz to $30$\,mW for $f=2.54$\,THz. The structures were placed in a temperature-regulated Oxford Cryomag optical cryostat, which for temperature $T=4.2$~K and below works as a He bath cryostat and for higher temperatures as a continuous flow cryostat. Samples were attached to an eight-pin socket and illuminated through $z$-cut crystal quartz windows. A thick black polyethylene film which is transparent in the terahertz range was mounted on the windows and prevented uncontrolled illumination of the sample with room light in both visible and infrared ranges. Off-axis parabolic mirrors were used to focus the laser radiation at the center of the sample. The laser beam shape at the position of the sample was measured with a pyroelectric camera.\cite{Ganichev1999} These measurements yielded the laser spot diameters (full width at half-maximum) of 1.5, 2.3, and 3.0~mm for laser lines with $f=2.54$, 1.62, and 0.69\,THz, respectively. Using these values, peak intensities were calculated. The radiation of the molecular gas laser is polarized linearly with the polarization axis depending on the particular laser mode.\cite{Ganichev2005} In some measurements we implement right- ($\sigma^+$) and left- ($\sigma^-$) handed circularly polarized radiation obtained using $\lambda/4$ plates made of $x-$cut crystal quartz. Laser radiation was modulated  at chopper frequency and photoresponse was measured applying standard lock-in technique, either as photocurrent $J_{x,y}$ or the corresponding photovoltage $U_{x,y} \propto J_{x,y}$ picked up across $50\, \Omega$ or $10$\,M$\Omega$ load resistors, respectively, see Fig.~\ref{fig1}(b). The photoresponse was mostly measured for the normal incidence of the THz radiation and in presence of a magnetic field $B$ up to 7~T applied  normal to the film surfaces. In some measurements specified below samples were rotated around the $y$-axis yielding simultaneously a variation of the normal magnetic field component and an oblique incidence of the THz radiation. Concurrent with the photogalvanic signal, we measured the radiation transmission using a pyroelectric cell detector placed behind the sample, see Fig.~\ref{fig1}(b).

\section{Experimental Results}

\begin{figure}[t]
\begin{center}
\includegraphics[width=\linewidth]{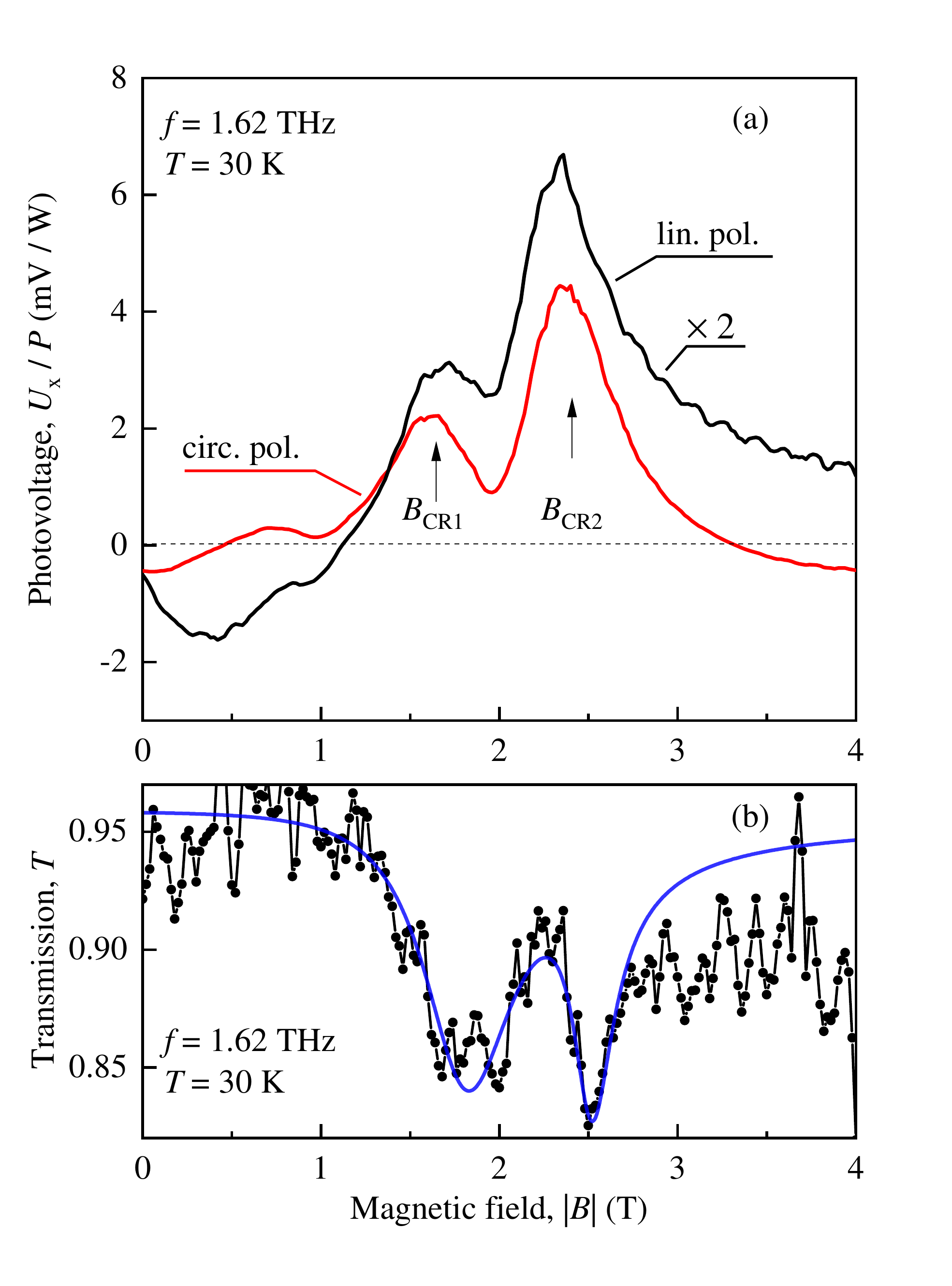}
\caption{ Magnetic field dependence of the photosignal $U_{x} \propto J_{x}$, panel (a), and of the simultaneously measured radiation transmission, panel (b). The data were obtained on the 200~nm HgTe film at $T=30$~K excited by linearly (black curve, circles) and circularly (red curve) polarized radiation of frequency $f= 1.62$~THz. Vertical arrows indicate positions of the resonant peaks. Solid line in panel (b) displays fit  after Eqs.~\eqref{T}, \eqref{tilde}, and \eqref{Gamma} obtained with fitting parameters $B_{\rm CR1}=1.8$~T and $B_{\rm CR2}=2.5~T$,  Fabry-Perot interference phase $\phi=8^\circ$, and with carrier densities and mobilities in top and bottom surface states  $n_1=1.0 \times 10^{11}$~cm$^{-2}$, $\mu_1=3.5 \times 10^{4}$~cm$^2$/V$\cdot$s, and $n_2=0.4\times 10^{11}$~cm$^{-2}$,  $\mu_2=8 \times 10^{4}$~cm$^2$/V$\cdot$s, respectively.
}
\label{fig3}
\vspace{-1cc}
\end{center}
\end{figure}

\begin{figure}[t]
	\begin{center}
		\includegraphics[width=\linewidth]{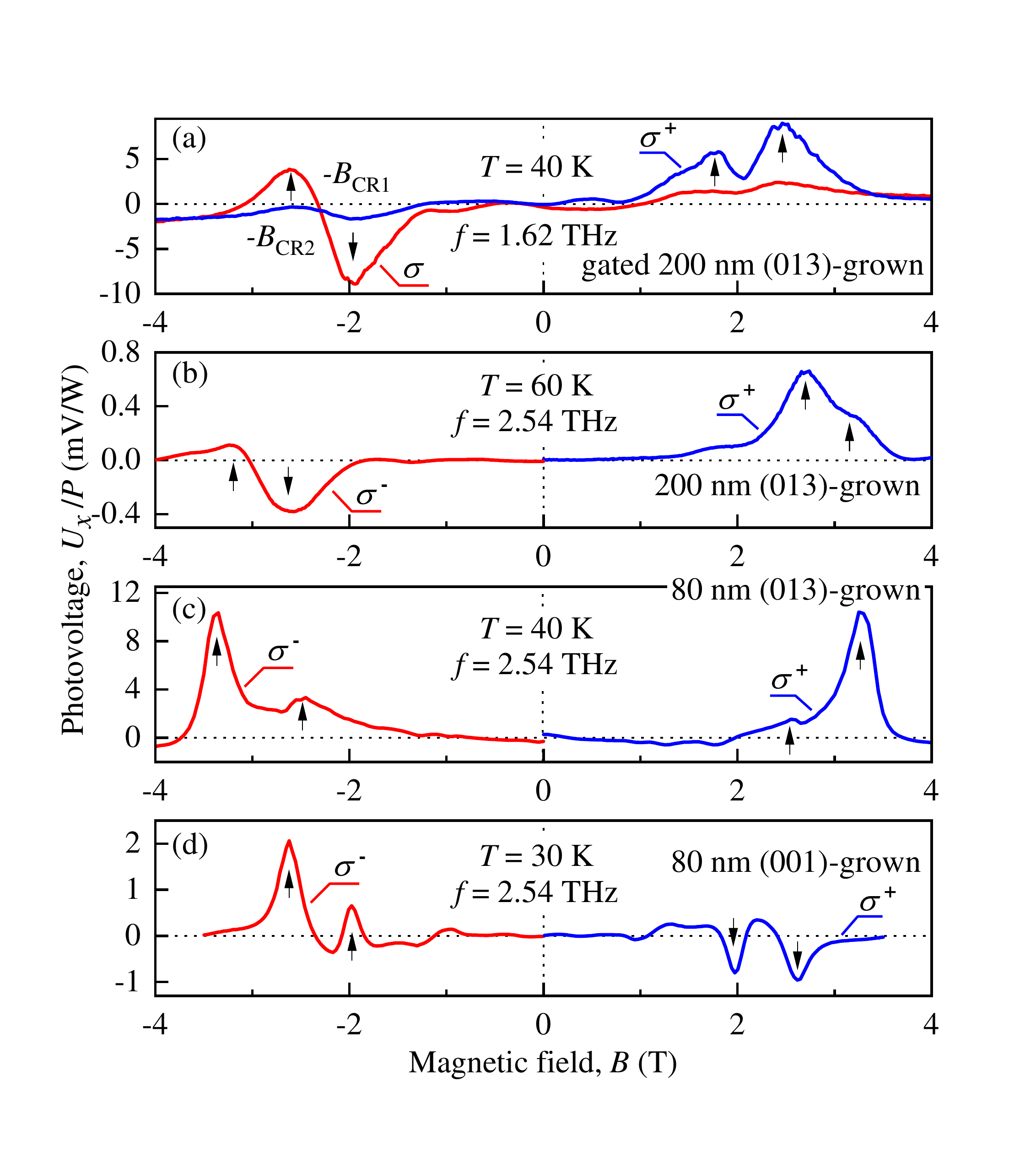}
		\caption{(a) Magnetic field dependencies of the photosignal measured in a 200~nm thick film for right- (blue curve) and left- (red curve) circularly polarized radiation. (b) - (d) Typical magnetic field dependencies of the photosignal measured in  films of different thicknesses at various temperatures for right- (blue curves) and left- (red curves) circularly polarized radiation. The figures demonstrate that photovoltage may display both even and odd behavior upon inversion of the magnetic field. }
		\vspace{-1cc}
		\label{fig4new}
	\end{center}
\end{figure}


\section{Experimental results on the photoresponse}

Figure~\ref{fig3}(a) shows the magnetic field dependence of the photosignal induced by illumination of a 200~nm HgTe film at $T=30$~K with terahertz radiation of $f = 1.62$~THz. The data were obtained for normal incidence and magnetic field applied perpendicularly to the film plane. Two resonant peaks of the photosignal are clearly detected at magnetic fields $B_{\rm CR1} = 1.6$~T and $B_{\rm CR2}=2.5$~T. While for linearly polarized radiation the peaks are detected for both magnetic field polarities, in the case of circularly polarized radiation they are only present for a single polarity (namely, positive $B$ for the right-handed circular polarization $\sigma^+$, and negative $B$ for the left-handed polarization $\sigma^-$), see Figs.~\ref{fig4new}(a) and \ref{fig4new}(b). \cite{footnote1} Figure~\ref{fig4new}(b) also reveals that the resonant photovoltage signals  $U_x\propto J_x$ can be either positive or negative, see discussion below. 
 
\begin{figure}[t]
	\begin{center}
		\includegraphics[width=\linewidth]{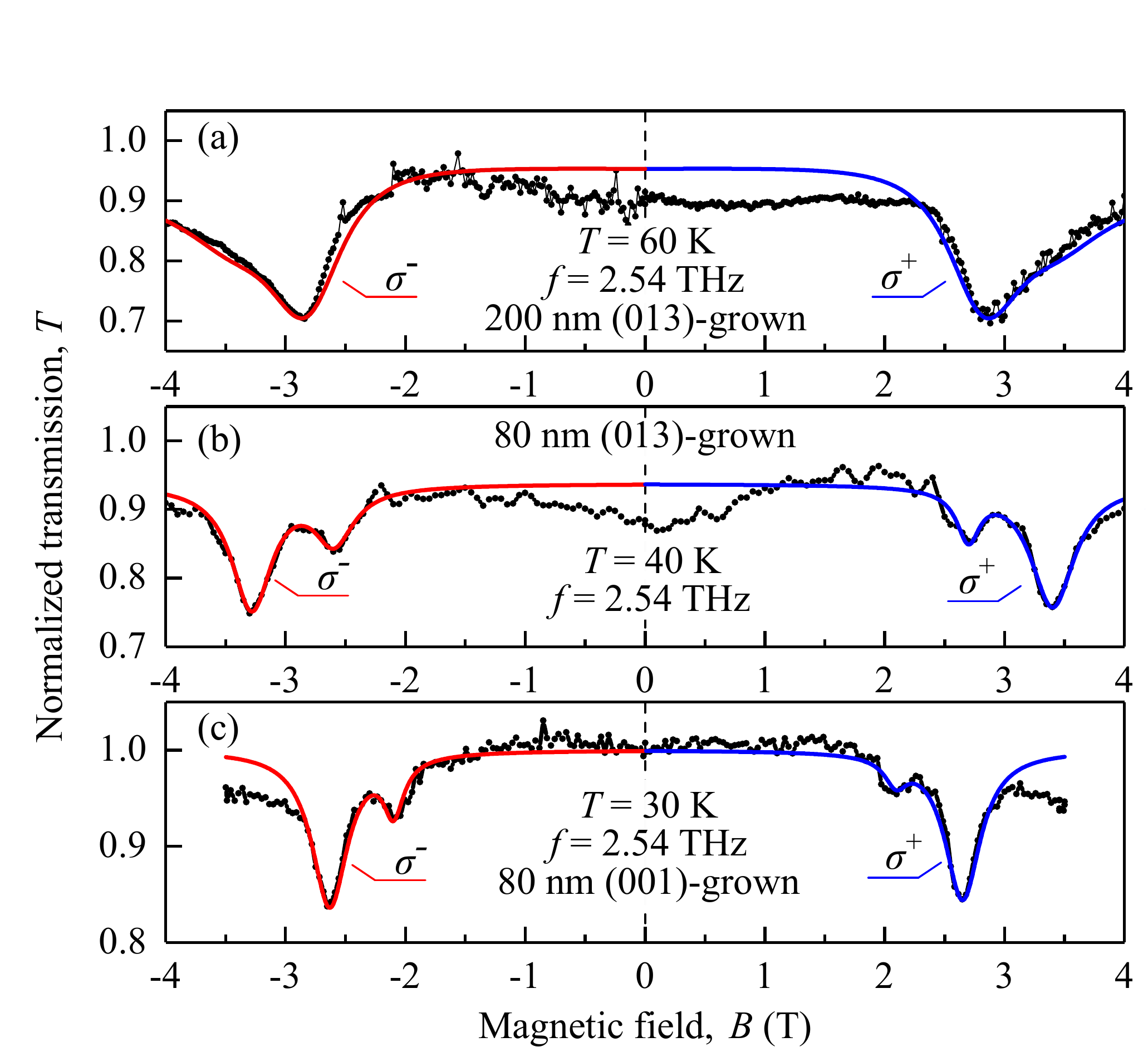}
		\caption{Magnetic field dependencies of the radiation transmission obtained for right-handed (positive $B$) and left-handed (negative $B$) circularly polarized radiation on HgTe films of different thicknesses. Solid lines show fits according to Eqs.~\eqref{T}, \eqref{tilde}, and \eqref{Gamma} obtained with fitting parameters $B_{\rm CR1}$ and $B_{\rm CR2}$,
			and the Fabry-Perot interference phase $\phi$, carrier densities ($n_1$ and $n_2$) and mobilities $\mu_1$ and $\mu_2$ in top and bottom surface states, respectively.  Parameters used are: panel (a) $B_{\rm CR1}=2.8$~T, $B_{\rm CR2}=3.4$~T, $\phi=12^\circ$, $n_1=2.5 \times 10^{11}$~cm$^{-2}$, $n_2=0.7\times 10^{11}$~cm$^{-2}$, $\mu_1=3 \times 10^{4}$~cm$^2$/V$\cdot$s, $\mu_2=2.3 \times 10^{4}$~cm$^2$/V$\cdot$s; panel (b) $B_{\rm CR1}=2.6$~T, $B_{\rm CR2}=3.3$~T, $\phi=0$, $n_1=0.8 \times 10^{11}$~cm$^{-2}$, $n_2=0.34\times 10^{11}$~cm$^{-2}$, $\mu_1=5.6 \times 10^{4}$~cm$^2$/V$\cdot$s, $\mu_2=5.3 \times 10^{4}$~cm$^2$/V$\cdot$s; and panel (c) $B_{\rm CR1}=2.1$~T, $B_{\rm CR2}=2.6$~T, $\phi=0$, $n_1=0.5 \times 10^{11}$~cm$^{-2}$, $n_2=0.1\times 10^{11}$~cm$^{-2}$, $\mu_1=6 \times 10^{4}$~cm$^2$/V$\cdot$s, $\mu_2=10 \times 10^{4}$~cm$^2$/V$\cdot$s.
%
%
	}
		\label{fig6}
		\vspace{-1cc}
	\end{center}
\end{figure}

\begin{figure}[t]
	\begin{center}
		\includegraphics[width=\linewidth]{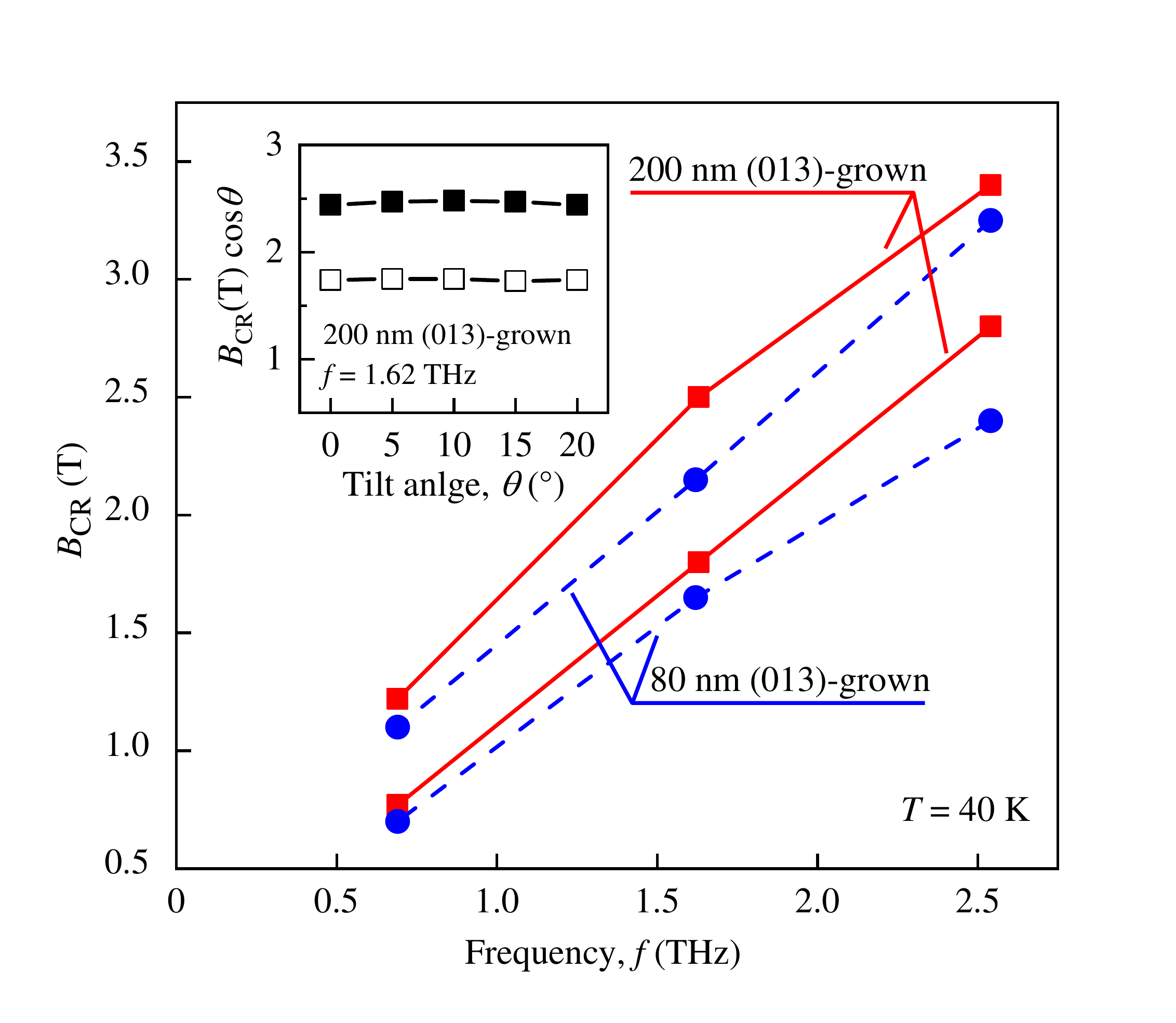}
		\caption{Frequency dependencies of the resonance fields $B_\text{CR1}$ and $B_\text{CR2}$ obtained from transmission measurements on 200~nm (squares) and 80~nm (circles) thick (013)-oriented films. The inset shows dependecies of $z$-components of $B_\text{CR1,2}$ on tilt angle 
			$\theta$. Here, the normal component of the CR	magnetic field, given by $B_{\rm CR1, CR2} \cos(\theta)$, is plotted against the tilting angle.
		}
		\label{fig4}
		
	\end{center}
\end{figure}

The resonant features with the same polarization dependence and at very close magnetic field positions are also detected in the radiation transmission measured simultaneously with the photosignal. These results are shown in Fig.~\ref{fig3}(b) together with a fit after Eqs.~\eqref{T}, \eqref{tilde}, and \eqref{Gamma} obtained with fitting parameters $B_{\rm CR1}$ and $B_{\rm CR2}$, carrier densities and mobilities in top and bottom surface states, and the Fabry-Perot interference phase $\phi$. Note that while in the photoresponse two peaks can always be resolved, in the transmission we observe that the resonant dips at high temperatures merge, see Fig.~\ref{fig6}(a).  Applying different laser lines we found that the resonance positions linearly scale with the radiation frequency $f$, see Fig.~\ref{fig4}.  All these facts reveal that the observed resonant  features in the photoresponse  and magnetotransmission are caused by the cyclotron resonance.
 
\begin{figure}[h]
	\begin{center}
		\includegraphics[width=\linewidth]{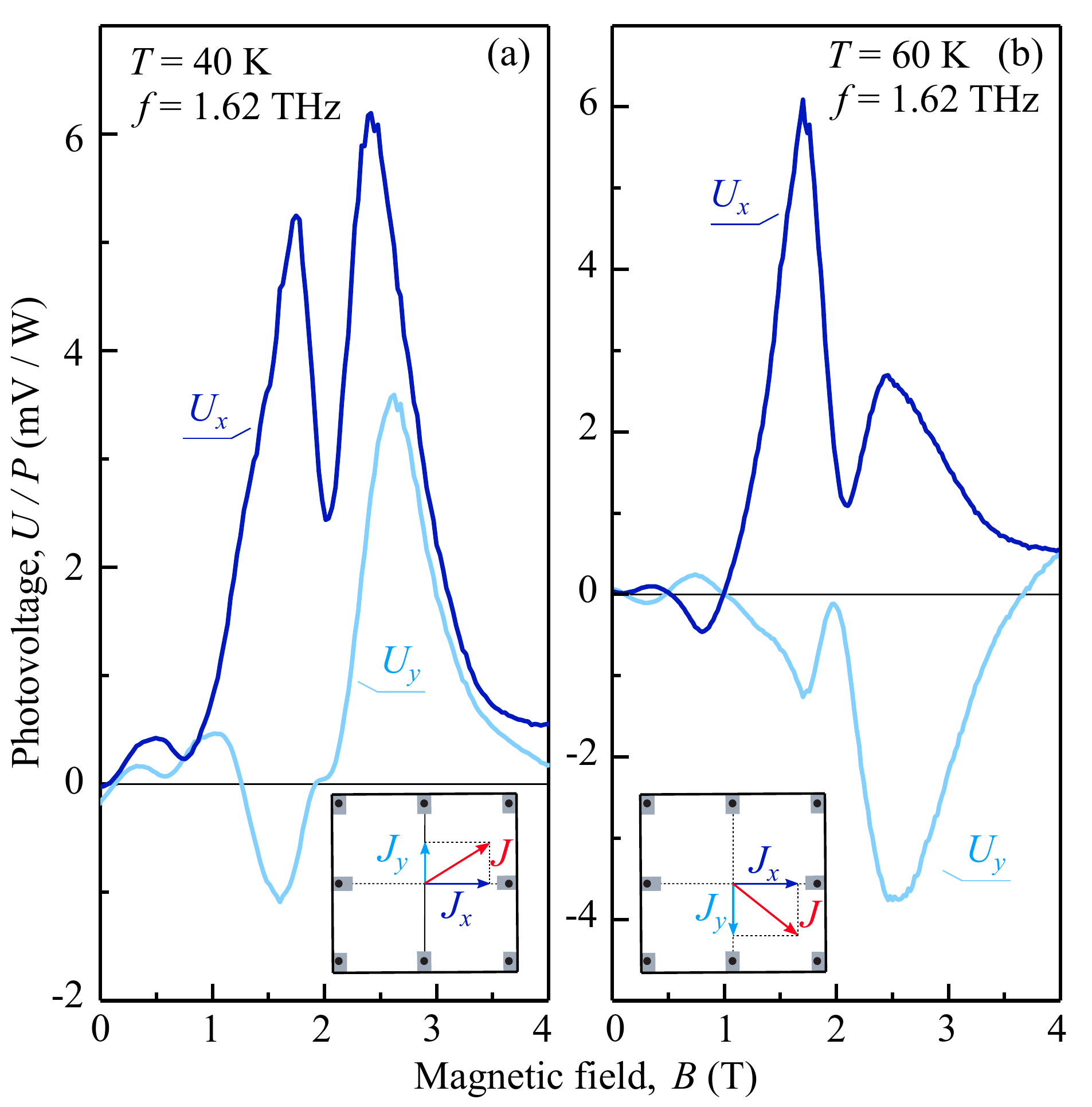}
		\caption{Typical magnetic field dependencies of the photosignals $U_x\propto J_x$ and $U_y \propto J_y$. The data in two panels were obtained on the 200~nm film at different temperatures.  Red thick arrows in insets sketch the direction of the total photocurrent. Horizontal and vertical arrows display the $x$- and $y$-projections of the  photocurrent  $\bm J$ measured in the experiment.
		}
		\label{fig7}
		\vspace{-1cc}
	\end{center}
\end{figure}

Depending on the experimental conditions, both positive and negative resonant photocurrents have been detected. This is seen in Fig.~\ref{fig4new}(b), where the resonant photocurrents at $\pm B_\text{CR1}$ have opposite signs, whereas the resonant photocurrents at $\pm B_\text{CR2}$ are positive for both polarities of the magnetic field. Furthermore, the sign of the resonant photocurrent can be either the same or opposite for the $x$- and $y$-projections of the total photocurrent. This is illustrated in Fig.~\ref{fig7} demonstrating resonant photosignals $U_x\propto J_x$ and $U_y \propto J_y$. Moreover, the sign of the photoresponse may change under variation of such experimental parameters as temperature. This is shown in Figs.~\ref{fig7}(a) and \ref{fig7}(b) for the resonances in $U_y \propto J_y$ at  $B_{\rm CR2}$ for temperatures 40 and 60~K. Note that similar variations of photocurrent projections on $x$ and $y$ directions with temperature and wavelength have been previously reported for (013)-grown samples studied in the absence of magnetic field.\cite{Wittmann2010}

To conclude on whether the observed resonances originate from carriers in the \textit{bulk} of the HgTe film or those occupying \textit{two-dimensional} surface states, we carried out measurements with the magnetic field tilted by an angle $\theta$. The inset in Fig.~\ref{fig4} shows that positions of the resonances are determined by the magnetic field component $B_z$, normal to the film surface. Indeed, calculating from the peak positions $B_{\rm CR1,2}$ the values of $B^{\rm{CR1,2}}_z = B_{\rm{CR1,2}} \cos(\theta)$, we obtain that both resonance positions $B^{\rm{CR1,2}}_z$ do not depend on the angle $\theta$. This finding clearly indicates the two-dimensional nature of carriers responsible for the observed resonant effects. Cyclotron masses calculated from the resonance positions are given in Tab.~\ref{tab:threecols}.

\begin{figure}[t]
	\begin{center}
		\includegraphics[width=\linewidth]{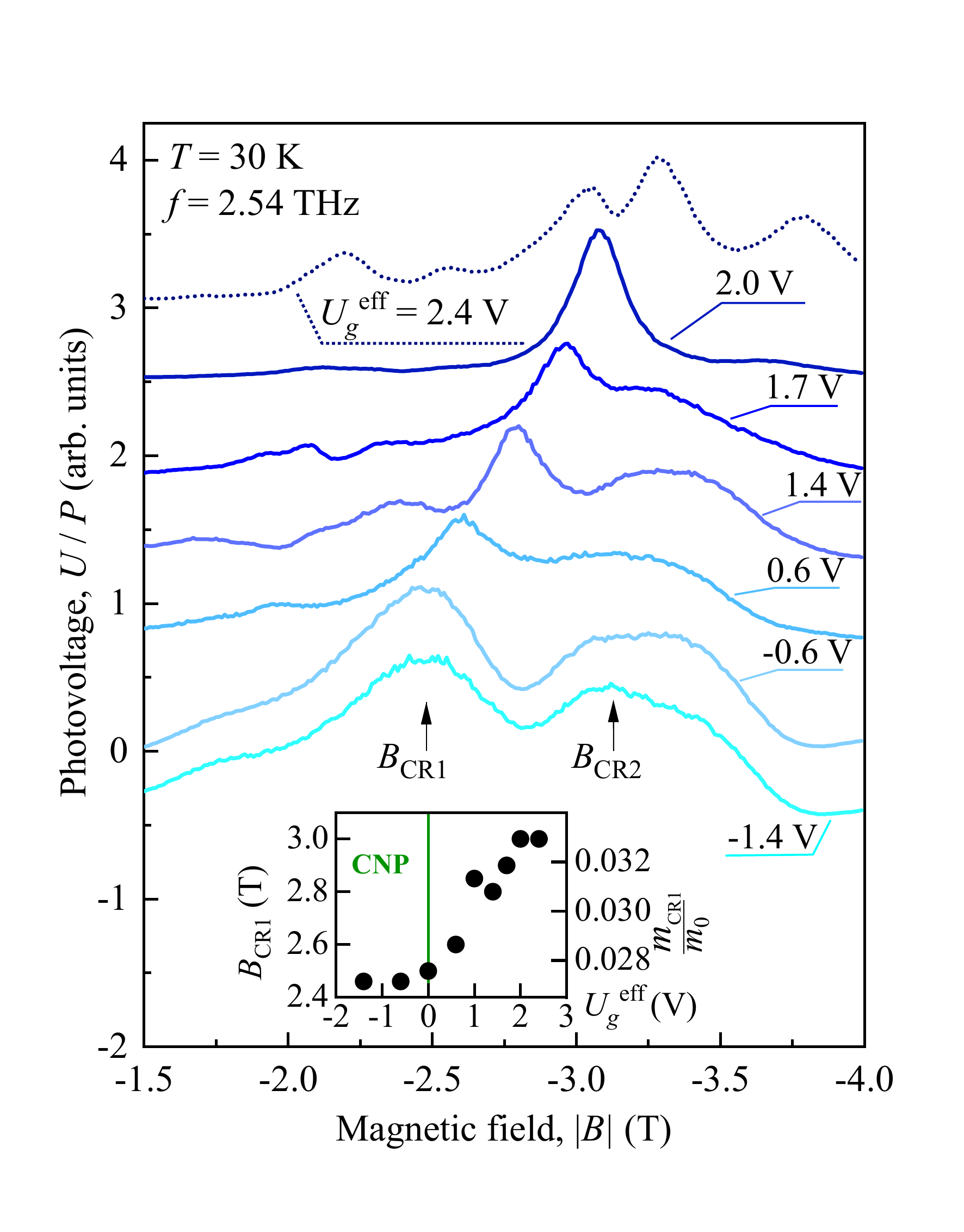}
		\caption{ Magnetic field dependencies of the photocurrent measured for the gated 200~nm thick sample at different effective gate voltages, $U_g^{\rm eff}=U_g-U_g^{\rm CNP}$, where $U_g^{\rm CNP}$ is the gate voltage corresponding to the charge neutrality point (vertical green line). Note that the data for different $U_g^{\rm eff}$ are $y$-shifted for clarity. Inset shows the gate voltage dependence of the $B_{\rm CR1}$ position. Right axis indicates the corresponding cyclotron mass $m_{\rm CR1}/m_0$. 
			}
		\vspace{-1cc}
		\label{fignew3}
	\end{center}
\end{figure}

In order to provide an additional support for the conclusion that the observed resonant photocurrent and transmission stem from two-dimensional surface states, we  studied the dependence of the CR assisted photocurrent on the Fermi energy position using gated structures with 200~nm HgTe layers, see Fig.~\ref{fignew3}. The data show that with increasing the gate voltage and, therefore, Fermi energy, the resonance at $B_{\rm CR1}$ shifts towards higher magnetic fields. Due to the linear dispersion, this behavior is typical for surface states. The position of the resonance at $B_{\rm CR2}$ is, however, almost independent of the gate voltage. The resonance at $B_{\rm CR2}$  is attributed to the bottom surface, whose electron density is not much influenced by the gate voltage. The same behavior has been observed in 80 nm thick films, both in CR (see Fig.~8 in Ref.~[\onlinecite{Dantscher2015}]) and magnetotransport measurements (see Fig.~1 in Ref.~[\onlinecite{Kozlov2014})], as well as in magnetotransport measurements on 200~nm films made from the same wafer, see Fig.~3 in Ref.~[\onlinecite{Savchenko2018}].  In particular, magnetotransport experiments show that indeed the filling rates $dN_s/dU_g$ for the top and bottom surface states are considerably different due to electrostatic screening of the bottom surface by the top one. From the resonance positions of the top surface state response we determined the gate voltage dependence of the cyclotron mass of the top state, see inset in Fig.~\ref{fignew3}.

Cyclotron resonance in transmission as well as the CR induced photocurrents are also detected in fully strained 80~nm HgTe films, which were previously shown\cite{Kozlov2014,Dantscher2015} to be fully gapped 3D topological insulators. Radiation transmission data measured in 80~nm thick films grown on (013)- and (001)-oriented substrates together with corresponding fits after Eqs.~\eqref{T}, \eqref{tilde}, and \eqref{Gamma} are shown in  Figs.~\ref{fig6}(b) and \ref{fig6}(c). Comparison of the resonant peak positions detected for 200 and 80~nm films is given in Tab.~\ref{tab:threecols}. It shows that in (013)-oriented films the positions of resonances are very close despite quite different thicknesses of the films. In previous study, see Ref.~[\onlinecite{Dantscher2015}], it has been shown that such two resonances originate from the topological surface states at the top and bottom intefaces of the 80~nm HgTe film. The coincidence of the resonance positions in the 200 and 80~nm films thus indicates that also in our thicker sample the resonances can be attributed to the top and bottom surface states. For the 80~nm films grown on (001)-surface, however, both resonances are detected at about 25\% lower magnetic fields. Like in 200~nm samples, the photocurrents measured in 80~nm films of both crystallographic orientations show helicity-sensitive resonances at magnetic fields corresponding to the CR positions detected in the radiation transmission, see Figs.~\ref{fig6}(b) and \ref{fig6}(c). Note that in 80~nm films we also observe different signs of the resonant responses for positive and negative magnetic fields, see for instance Fig.~\ref{fig6}(c), as well as for different contact pairs or temperatures.

\begin{table}
	\centering
		\begin{tabular}{r|c|c|c|c}
			 \toprule
			Sample	& $B_{\mathrm{CR1}}$, T & $B_{\mathrm{CR2}}$, T & $m_{\mathrm{CR1}}/m_0$ \, & $m_{\mathrm{CR2}}/m_0 $ \\
			\midrule
		200 nm (013) & 1.8 & 2.5 & 0.031 & 0.043   \\
		\,\,80 nm (013) & 1.6 & 2.2 & 0.028 & 0.038 \\
		\,\,80 nm (001) & 1.3 & 1.7 & 0.023 & 0.030 \\
	\bottomrule
	\end{tabular}
\caption{Cyclotron resonance positions and corresponding cyclotron masses extracted from transmission measurements at $T= 40$\,K using $f= 1.62$\,THz radiation on the 200~nm thick HgTe film grown on (013)-oriented substrate, and on 80~nm thick HgTe films grown on (013)- and (001)-oriented substrates.}
	\label{tab:threecols}
\end{table}

\vspace{-1cc}

\section{Discussion and theory of cyclotron resonance assisted photogalvanic current}

Below we analyse the data on the CR response of the surface states in transmission and photocurrent and possible reasons for partial strain remaining in 200~nm thick HgTe films. We discuss fits used for the data on the radiation transmission,  address the origin of the photocurrent excited in (013)-oriented films, explain the observed difference in the sign of the CR induced photocurrents,  and discuss the surprising observation that CR assisted photocurrents in (001)-grown films have been excited by \textit{normally} incident radiation and for magnetic fields applied perpendicularly to the film surface.  

\subsection{Surface states and strain relaxation}

The data described above  demonstrate that, in spite of the fact that  the thickness of 200~nm films is slightly larger than the estimated critical thickness of lattice relaxation, the surface states with characteristics very close to those of the  fully strained 80\,nm 3D TI films  can clearly be detected. This conclusion is supported by: (i) two-dimensional origin of the CR (Fig.~\ref{fig4}, inset), (ii) presence of two CRs corresponding to excitation of the bottom and top surfaces  (Figs.~\ref{fig3}, \ref{fig4new}, \ref{fig4}, \ref{fig7}), and (iii) very close values of cyclotron masses obtained from both resonances in (013)-grown 200 and 80~nm thick films  (Fig.~\ref{fig4}). Our results are in agreement with magnetotransport and $X$-ray data, see Ref.~[\onlinecite{Savchenko2018}], obtained on samples made from the same wafers as samples studied in the present work. 
An additional direct support of our conclusion on the detection of resonances excited in topological surface states can be found in Ref.~[\onlinecite{Savchenko2018}]. In this work, a 200 nm HgTe film made from the same wafer as here was investigated by means of transport measurements combined with capacitance spectroscopy. The detailed analysis of Shubnikov-de Haas oscillations both in conductivity and capacitance allowed to unambiguously distinguish three groups of electrons identified as bulk electrons and electrons on the top and bottom surfaces. It was established that the surface states are spin non-degenerate, while the bulk electrons are spin-degenerate.

The $X$-ray diffraction  measurements demonstrated that the 200~nm thick film still has about 60\% strain (100\% means that HgTe film is fully strained and has an indirect bulk band gap of the order of 15~meV,\cite{Brune2011,Kozlov2016,Dantscher2015} 0\% means that it is fully relaxed to its own lattice constant and has few meV of band overlap).  Let us discuss possible reasons of the considerable remaining  strain in films with substantially larger thickness as compared to the estimated critical thickness reported previously. Calculations of the critical thickness of a strained epitaxial film, above which the formation of misfit dislocations reducing the strain becomes energetically favourable, usually rely on the Matthews model.\cite{Matthews1975} This model employs an equilibrium thermodynamic approach, and thus neglects kinetic factors related to dislocation formation at a finite rate of epitaxial growth. The Matthews approximation demonstrates good agreement with experimental data in the case of large mismatch of the lattice parameters between the epilayer and substrate, but is less reliable for small misfits typical for HgTe/CdTe heterostructures. Furthermore, experimental data for low-strain semiconductor systems demonstrate that the mechanism of  strain relaxation is not sufficiently well described by existing models.\cite{Pichaud2002,France2011a,France2011} For instance, a discrepancy reaching two orders of magnitude between observed and calculated values of the critical thickness was reported for low-strain Ge$_x$Si$_{1-x}$/Si.\cite{Bolkhovityanov2001}

 A further reason is the substrate orientation. Theoretical values for the critical thickness of HgTe on (001)-oriented CdTe (lattice mismatch of about $0.3\%$) range between 50~nm, see Ref.~[\onlinecite{Basson1983}], and 200~nm, see Refs.~[\onlinecite{Brune2011,Ballet2014,Berding2000}]. In the case of Hg$_{0.7}$Cd$_{0.3}$Te grown on (013)-oriented CdTe like in samples studied here, the lattice mismatch is $0,216\%$ and the calculated critical thickness varies from 47.5 to 344~nm for 12 possible slip systems of dislocations classified by orientation of the slip plane and Burgers vector.\cite{Sidorov2015} The experimental part of Ref.~[\onlinecite{Sidorov2015}] reports that dislocations begin to form when the film thickness exceeds 80~nm. Starting from 150~nm, a dislocation network emerges which results in about 40\% strain relaxation. Similar results on relaxation in HgTe on (001)-oriented CdTe are reported in Ref.~[\onlinecite{Berding2000}].

 At last, but not least, the situation becomes even more complex in multilayer heterostructures grown on substrates with large mismatch, such as HgTe films grown on GaAs substrate with additional ZnTe and CdTe layers (CdTe/ZnTe/GaAs). In this case it is necessary to take into account significantly different thermal expansion coefficients of GaAs and epitaxial layers. In the presence of this additional factor, one can expect that strain remains strong even for heterostructures containing thick HgTe films, especially after the sample is cooled down for low-temperature measurements. 

Summarizing, the observation that considerable strain remains in HgTe films with thickness larger than the critical thickness calculated in the framework of the Matthews model is attributed to the necessity to extend the model by (i) considering of the dislocation network formation and/or additional relaxation mechanism in the HgTe/CdTe system, (ii) crystallographic orientation of the substrate, as well as (iii) different thermal expansion coefficients of GaAs substrate and epitaxial layers.
  
 \subsection{Transmission fits}
 
 Our analysis of the transmission data follows Ref.~[\onlinecite{Herrmann2016}] and is similar to that presented in Refs.~[\onlinecite{Abstreiter1976,Chiu1976,Mikhailov2004,Zhang2014}]. We consider a circularly polarized wave with frequency $\omega/2\pi$ normally incident upon a sample which is modeled as a dielectric slab of thickness $w$ (it represents the thick GaAs substrate having the refraction index $n=3.6$) with an infinitely thin conducting layer on top of it (characterized by the total dynamic conductivity combining contributions of surface and bulk carriers of the conducting HgTe layer). This model takes into account that the whole MBE-grown heterostructure containing the HgTe layer has a thickness much smaller than the wavelength and, in this sense, is effectively two-dimensional. By contrast, the optical phase shift $\phi=n \omega w/c$ across the substrate is large, $\phi\gg 1$, which necessitates an account for multiple reflections between the dielectric interfaces (Fabry-Perot interference). The resulting expression for the power transmission has the following form:\cite{Herrmann2016,Abstreiter1976}
 \begin{equation}\label{T}
 	T=\left|(1+\tilde{\sigma})\cos\phi-i\dfrac{1+n^2+2\tilde{\sigma}}{2n}\sin\phi\right|^{-2}.
 \end{equation}
 Here the dimensionless $\tilde{\sigma}=\sigma/2\epsilon_0 c$ represents the complex dynamic conductivity $\sigma=\sigma_{xx}(\omega)+ i\sigma_{xy}(\omega)$ (we assume isotropic transport in the HgTe layer, $\sigma_{xx}=\sigma_{yy}$ and $\sigma_{yx}=-\sigma_{xy}$); $\epsilon_0$ is the permittivity of free space. Using the Drude expression for the ac conductivity, one obtains 
 \begin{equation}\label{tilde}
 	\tilde{\sigma}=\dfrac{\Gamma}{\gamma-i(1+\omega_c/\omega)}.
 \end{equation}
 Here $\omega_c=eB/m$ is the cyclotron frequency. The dimensionless parameter
 \begin{equation}
 	\label{Gamma}
 	\Gamma=\dfrac{e^2 n_e}{2 \epsilon_0 c m\omega}=\dfrac{0.301\ \text{T}}{B_\text{CR}}\ \dfrac{n_e}{10^{12}\ \text{cm}^{-2}}
 \end{equation}
 represents the so called radiative or superradiant decay.\cite{Abstreiter1976,Chiu1976,Mikhailov2004,Zhang2014} It is fully determined by the electron density $n_e$ and the position of the CR $B_\text{CR}=\omega m/e$. The dimensionless parameter
 \begin{equation}
 	\label{gamma}
 	\gamma=\dfrac{1}{\omega\tau}=\dfrac{1}{\mu B_\text{CR}}
 \end{equation}
 represents the transport relaxation time $\tau$ and corresponding mobility $\mu$. Expressions above correspond to the left-handed circular polarization (CR at $\omega_c=-\omega$), while those for the right-handed polarization are obtained via the substitution $\omega_c\to-\omega_c$. In the case of several carrier types, $\tilde{\sigma}$ is given by the sum of conductivities in all transport channels. 
 Therefore, apart from the unknown Fabry-Perot interference phase $\phi$, the $B$-dependence of the transmitted power is determined by the cyclotron masses, concentrations, and mobilities of all individual components. The fit parameters used in the calculations are provided in captions to Figs.~\ref{fig3} and \ref{fig6}. It is seen that our transmission data can indeed be well described by the above theory. The fit parameters are in agreement with recent transport measurements (see Ref.~[\onlinecite{Savchenko2018}]) which allow one to extract basic transport quantities for individual layers and the bulk. In particular, in transmission fits we obtained the electron density in top surface states about 3 times larger than that in the bottom states, in full agreement with more recent results of Ref.~[\onlinecite{Savchenko2018}], see Fig. 3 there.
  Due to Fabry-Perot interference the form of the CR dips in magnetotransmission is sometimes found rather asymmetric, see e.g. Fig.~\ref{fig6}(a). The usual symmetric Lorentzian form is recovered only for $\phi/\pi$ close to integer values (constructive interference) or half-integer values (destructive interference) provided that the CR features originating from different transport channels are well separated. The broadening of the CR is affected by both $\gamma$ and $\Gamma$, which in our case are comparable in magnitude. The transport data in Fig.~\ref{fig2} suggest that in the studied temperature range mobility (and, therefore, $\gamma$) stays constant, while the density ($\Gamma$) increases by an order of magnitude. This is consistent with our observation of two separate sharp CR dips in transmission in Fig.~\ref{fig3}(b) at $T=30$~K and their strong broadening and merging at higher $T=60$~K in Fig.~\ref{fig6}(a).
 
\subsection{PGE theory}

 Cyclotron resonance assisted photocurrents have been previously detected in different materials including InSb/InAlSb quantum wells (QWs),\cite{Stachel2014} 3D TI HgTe strained films,\cite{Dantscher2015} and HgTe QWs of different thicknesses including the critical thickness.\cite{Olbrich2013} Despite similar manifestation and phenomenological description, the microscopic origin of the effect is not universal. The underlying microscopic mechanisms can involve both spin and orbital degrees of freedom, and are sensitive to details of the band structure. A theoretical model of the orbital CR-induced photogalvanic currents in 3D TI HgTe strained films has been introduced in Ref.~[\onlinecite{Dantscher2015}]. It describes $B$-dependent photocurrents emerging under a uniform illumination in systems with linear dispersion lacking spatial inversion symmetry. A detailed theory of the linear photogalvanic effect caused by magneto-induced asymmetry under photoexcitation for asymmetric QWs and for surface states of bulk TIs was developed in Refs.~[\onlinecite{budkin2016,budkinnano2015}]. In particular, the theory predicts a resonant enhancement of the photocurrents in the vicinity of CR. Here we present a theory of the photocurrent generated in the process of energy relaxation of photo-excited carriers, which explains all major experimental findings presented above. The key ingredient of the model is an asymmetry of electron-phonon scattering rate in the momentum space which emerges in the presence of an applied magnetic field.

In thick films, as discussed in the present work, the penetration depth of wavefunctions in both surface states is much smaller than the film width. Thus the coupling between surface states is negligibly small, which allows us to treat them independently. At each of the interfaces, we consider the process of energy relaxation of a spin-nondegenerate two-dimensional electron gas heated by the THz radiation. 	It is assumed that at experimentally relevant low temperatures the momentum relaxation of carriers in the surface states is dominated by elastic scattering off static defects, while the energy relaxation of photo-excited carriers is determined by their interaction with acoustic phonons.

Under continuous illumination, electrons partially absorb the energy of THz radiation and transfer the absorbed energy to the phonon bath. This energy transfer is naturally accompanied by heating of the electron subsystem. We assume that frequent electron-electron collisions lead to fast exchange of energy leading to thermalization of the photo-excited carriers. 	As a result, the isotropic part of the steady-state nonequilibrium distribution function has a Fermi-Dirac form, $f^{(0)}_k=1/\{\exp[(\varepsilon_k-\varepsilon_F)/T_e]+1\}$. 	Here $k$ and $\varepsilon_k$ denote the momentum and energy spectrum of electrons, $\varepsilon_F$ is the chemical potential, and $T_e=T+\Delta T$ is the effective electron temperature, which is higher than the temperature of the phonon bath $T$.

The asymmetry of electron-phonon scattering rate in $k$-space leads to emergence of an anisotropic part $\delta f_k$ of the steady-state nonequilibrium distribution function, which can be found from the Boltzmann equation,
	\begin{equation}
	\label{kinetic}
	\dfrac{\delta f_k}{\tau}+\dfrac{e}{\hbar}[\bm{v}_k\times \bm{B}]\dfrac{\partial \delta f_k}{\partial \bm{k} }=
	St^{(ph)}[f^{(0)}_k].
	\end{equation}
Here the first term describes momentum relaxation ($\tau$ is the transport relaxation time), the second term  describes cyclotron motion ($\bm{v}_k$ is the group velocity), and $St^{(ph)}[f^{(0)}_k]$ is the electron-phonon collision integral. 	In the latter, we replaced the full distribution function with $f^{(0)}_k$. This corresponds to the leading first-order perturbation theory for the anisotropic correction $\delta f_k$ with respect to a weak asymmetry of $St^{(ph)}$.

The electron-phonon collision integral is given by\cite{gantmakher2012carrier}
	\begin{multline}
	St^{(ph)}[f^{(0)}]=
	\sum_{\bm{k'},\bm{q},\pm}
	W_{k'k}^{\pm}(\bm{q}) f^{(0)}_{k'}(1-f^{(0)}_{k})N_q^\pm\\
	-W_{k'k}^{\pm}(\bm{q}) f^{(0)}_k(1-f^{(0)}_{k'}) N_q^{\pm}\:,\nonumber
	\end{multline}
where $W^{\pm}_{\bm{k}'\bm{k}} (\bm{q}) =\dfrac{2\pi}{\hbar}|M_{\bm{k}'\bm{k}}^{\pm}(\bm{q})|^2  \delta(\varepsilon_{k'}-\varepsilon_{k}\pm\hbar \Omega_q) $ is the scattering probability, $\bm{q}$ is the phonon wave vector, $N_q^-$ is the phonon occupation number, $N_q^+=N_q^-+1$, and $\pm$ correspond to phonon emission/absorption. The matrix elements $M_{\bm{k}'\bm{k}}^{\pm}(\bm{q})$ are Hermitian conjugates, $M_{\bm{k}'\bm{k}}^{-}(\bm{q})=[M_{\bm{k}\bm{k}'}^{+}(\bm{q})]^*$.
In the relevant $T$- and $q$-range, the occupation numbers for acoustic phonons are large, $N_q^-\simeq T/\hbar \Omega_q\gg 1$. The anisotropic correction $\delta f_k$ found from Eq.~\eqref{kinetic} yields the electric current $\bm{j}=e \sum \bm{v}_k f_k$ given by
	\begin{multline}
	\label{current_eq}
	\bm{j}=e\dfrac{\tau}{1+\omega_c^2 \tau^2}\sum_{\bm{k},\bm{k}',\bm{q}}
	\{(\bm{v}_{k}-\bm{v}_{k'})+\omega_c \tau [(\bm{v}_{k}-\bm{v}_{k'})\times \bm{o}_z]\}\times\\
	W_{k'k}^{-}(\bm{q})
	[
	f^{(0)}_{k'} (1-f^{(0)}_{k})N_q^+
	-f^{(0)}_k (1-f^{(0)}_{k'})N_q^-
	]\:.
	\end{multline}
Here $\omega_c=e B_z/ m$ is the cyclotron frequency, and $\bm{o}_z$  is the unit vector normal to the surface. Assuming $\Delta T \ll T$, we linearize Eq.~\eqref{current_eq} with respect to $\Delta T$ using the relation $f^{(0)}_{k'} (1-f^{(0)}_{k})N_q^+ -f^{(0)}_k (1-f^{(0)}_{k'})N_q^-\approx f^{(0)}_{k'}(1-f^{(0)}_{k}) \Delta T/T$. In turn, the temperature difference $\Delta T$ can be related to the intensity of radiation $I$ from the energy balance equation
	\[
	I \eta(\omega)=\sum_{\bm{k}} \varepsilon_k St^{(ph)}[f^{(0)}_k]\:,
	\]
where $\eta(\omega)$ is the absorbance.

In the absence of a static magnetic field the current \eqref{current_eq} vanishes due to the time reversal symmetry (TRS).	In the presense of magnetic field the TRS operation involves the inversion of the magnetic field orientation, 	${W^{-}_{\bm{k}',\bm{k}} (\bm{q},\bm{B})=W^{+}_{-\bm{k},-\bm{k}'} (-\bm{q},-\bm{B})}$. As a result, 	magneto-induced asymmetry of electron scattering by phonons becomes possible, with the lowest order correction being proportional to $B_z$,
	\begin{equation}
	\label{probability_anisotropic}
	|M_{\bm{k}'\bm{k}}^{-}(\bm{q})|^2=\delta_{\bm{k}',\bm{k}+\bm{q}_{\parallel}}[w_0+w_1 B_z \bm{g}\cdot (\bm{k}'+\bm{k})]\:.
	\end{equation}
	
Here the Kronecker delta $\delta_{\bm{k}',\bm{k}+\bm{q}_{\parallel}}$ expresses the momentum conservation law, $\bm{q}_{\parallel}$ is the in-plane component of the phonon momentum, and $\bm{g}$ is the in-plane vector which defines the scattering  anisotropy. Phenomenologically, this type of scattering asymmetry is similar to that obtained in Ref.~[\onlinecite{tarasenko2008}] for quantum wells or in Refs.~[\onlinecite{averkiev2002,tarasenko2005,ivchenko2004}] for the spin-dependent scattering. A specific property of structures with (013) surface orientation is their low in-plane symmetry described by the $C_\text{1}$ point group. In this case, symmetry imposes no restrictions on the orientation of the two-dimensional vector $\bm{g}$.

Taking into account that the out-of-plane component of phonons is typically large, $ q_z \gg |\bm{k}-\bm{k}'|=q_{\|}$, one can perform an explicit summation over $\bm{k}$, $\bm{k'}$ and $\bm{q}$ in Eq.~\ref{current_eq} with scattering probability~\eqref{probability_anisotropic}. The resulting expression for the photocurrent reads
	\begin{equation}
	\label{final_current}
	\bm{j}=-e I \eta(\omega) B_z 
	\dfrac{\bm{g}+\omega_c \tau \bm{g}_{\perp} }{1+\omega_c^2 \tau^2}
	\xi 
	\dfrac{\langle |q_z| w_1 \rangle}{\langle |q_z| w_0 \rangle}\:.
	\end{equation}
Here $\bm{g}_{\perp}=[\bm{g} \times \bm{o}_z]$ is the in-plane vector perpendicular to the vector $\bm{g}$, triangle brackets stand for averaging over $q_z$ and directions of $\bm{k}$ and $\bm{k'}$. The parameter $\xi=1+m  \partial^2 \varepsilon_p/\partial p^2$ reduces to a numerical factor  $\xi=1(2)$ for the linear (parabolic) spectrum of the charge carriers. We note that all functions here are defined at the Fermi level.

Due to the CR peak in the absorbance $\eta(\omega)$, the photocurrent in Eq.~\eqref{final_current} is resonantly enhanced when the radiation frequency $\omega$ matches the cyclotron frequency $\omega_c$, thus reproducing the behavior observed in all studied samples, see Figs.~\ref{fig3}\,$-$\,\ref{fig4}. As expected, for circularly polarized radiation peaks are present only for one magnetic field polarity, whereas for linearly polarized radiation the resonant enhancement of the photocurrent is observed for both magnetic field polarities. The CR position is defined by the band structure and is different for top and bottom surfaces, see Tab.~\ref{tab:threecols}. The cyclotron masses obtained from magnetic fields $B_{\mathrm{CR1}}$ and $B_{\mathrm{CR2}}$  agree well with the estimated cyclotron masses of surface states obtained by $\bm{k} \cdot \bm{p}$ calculations of the band structure\cite{Dantscher2015}  adapting the procedure described in the supplementary material to Ref.~[\onlinecite{Brune2011}] and Refs.~[\onlinecite{Zholudev2012, Novik2005}]. 

According to Eq.~\eqref{final_current}, the projections of the generated total photocurrent on the $x$ and $y$ directions are given by 
 \begin{align}
 \label{projection_photocurrent}
 j_x=-e I \eta(\omega) B_z 
 \dfrac{g_x+\omega_c \tau g_y }{1+\omega_c^2 \tau^2}
 \xi 
 \dfrac{\langle |q_z| w_1 \rangle}{\langle |q_z| w_0 \rangle}\:,\nonumber\\
 j_y=-e I \eta(\omega) B_z 
 \dfrac{g_y-\omega_c \tau g_x }{1+\omega_c^2 \tau^2}
 \xi 
 \dfrac{\langle |q_z| w_1 \rangle}{\langle |q_z| w_0 \rangle}\:.
 \end{align}
While the magnitude of generated photocurrent is proportional to $|\bm{g}|$, its direction depends both on the direction of $\bm{g}$ and on the value of parameter $\omega_c \tau$. Moreover, in (013)-grown structures possessing no non-trivial in-plane point symmetry operations, the orientation of $\bm{g}$ is not bound to any crystallographic direction and thus may depend on temperature, gate voltage, carrier density etc., see Ref.~[\onlinecite{Wittmann2010}]. This can explain the observed sign changes of the resonant photocurrent, both under reversal of the magnetic field polarity and due to variations of other experimental parameters such as temperature. A clear indication of such changes is presented in Fig.~\ref{fig7} showing that the $y$-projection of the resonant photocurrent at $B_{\mathrm{CR2}}$ changes its sign with temperature. Note that in previous work on zero-$B$ photogalvanic currents in (013)-grown HgTe QWs a change of sign of the photogalvanic current projection with increase of temperature has also been detected for certain projections of the total current.\cite{Wittmann2010} Equations~\eqref{projection_photocurrent} show that the current projections $j_x$ and $j_y$ can be both even or odd functions of $B_z$, depending on the relative orientation of the vector $\bm{g}$ and the axis along which the photocurrent is measured. This is also detected in our experiments, see Fig.~\ref{fig4new}.

\subsection{Photocurrents in (001)-grown films}
 
We now turn to the discussion of our findings on the $80$\,nm HgTe film grown in (001) orientation, see Fig.~\ref{fig4new}(d).
Our samples are grown on GaAs substrates. There are several reasons for that. First, we have access to high-quality GaAs wafers only and not to CdTe ones. Second, the lattice mismatch between GaAs and CdTe is too high (13.6\%) for a defect-free MBE growth. In order to reduce the lattice mismatch, we use a thin ZnTe layer having the lattice constant in between of CdTe and GaAs. The details can be found in Ref.~[\onlinecite{Sidorov2001}]. Third, the (013) growth direction was experimentally found to be the optimal one, giving the maximum crystalline quality. The growth on the vicinal surface with (013) orientation reduces the dissociation energy of two-atomic Te molecules. The details of this process can be found in Ref.~[\onlinecite{Varavin1995}].

\begin{figure}[t]
 		\begin{center}
 			\includegraphics[width=\linewidth]{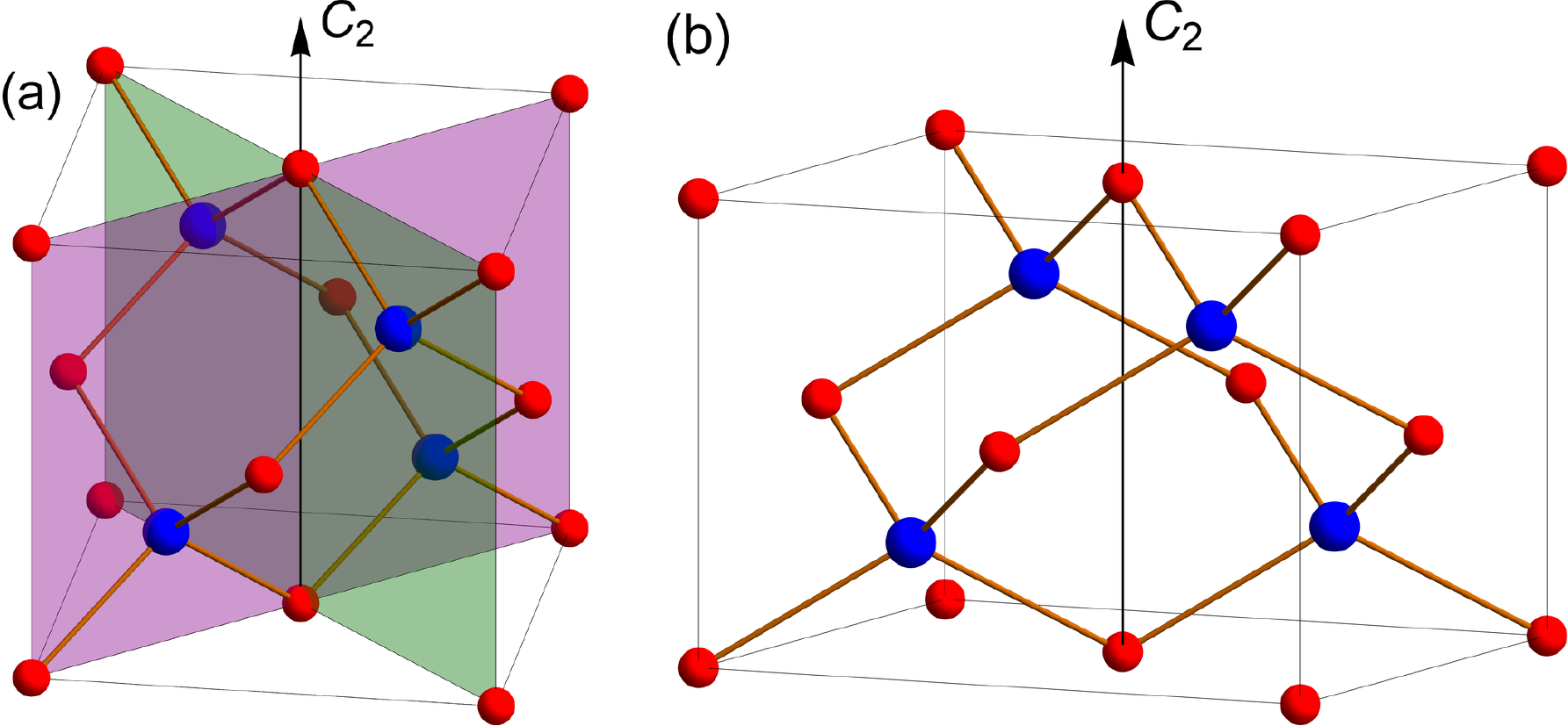}
 			\caption{(a) Schematic representation of symmetry operations in HgTe without strain: Two mirror planes, shown by semi-transparent colors, and one two-fold rotation axis. (b) Deformed unit cell of HgTe: The two-fold rotation axis remains a symmetry element of the considered system.}
 			\label{fig8}
 			
 		\end{center}
 	\end{figure}

The fact that the photocurrent, either resonant or nonresonant, can be excited at normal incidence and for a magnetic field perpendicular to the film surface is surprising. Indeed, bulk mercury telluride has zinc-blende structure which has a $T_\text{d}$ point symmetry group. The interface between the HgTe film and the CdHgTe cap layer in (001)-grown structure makes $z$ and $-z$ directions non-equivalent, which reduces the symmetry of the system from $T_\text{d}$ down to $C_\text{2v}$, with one two-fold rotation axis and two mirror planes shown in Fig.~\ref{fig8}(a). In systems that have a two-fold rotation axis photocurrents can be observed only for oblique incidence of radiation when both the in-plane and normal components of the radiation electric field are present.  Indeed, the general phenomenological expression for the photocurrent reads\cite{Belkov2005,Ivchenkobook,IvchenkoGanichev2017}	
 	\begin{equation}
 	\label{photocurrent_phemenology}
 	j_{\lambda}=\sum\limits_{\mu\nu}\chi_{\lambda \mu \nu}(\bm{B}) E_\mu E_\nu^*\:,
 	\end{equation}
with a third rank tensor $\chi$ depending on the static magnetic field $\bm{B}$. For the particular mechanism of photocurrent considered above, the tensor $\chi$ is proportional to the absorbance $\eta(\omega)$, see Eqs.~\eqref{final_current} and \eqref{projection_photocurrent}. In our experiments, the magnetic field is applied along the growth direction, $\bm{B}=(0,0,B_z)$; the THz electric field is polarized in the film plane, such that all indices $\lambda$, $\mu$, $\nu$ belong to the $(x, y)$-plane. Under $C_2$ rotation in the $(x, y)$-plane, $B_z$ and $\chi$ remain intact, while the in-plane current $j_\lambda$ and components of THz electric field $E_\mu$ and $E_\nu^*$ change sign. Thus the left-hand side $j_\lambda$ in Eq.~\eqref{photocurrent_phemenology} transforms to $-j_\lambda$, while the right-hand side remains unaltered. The $C_\text{2}$ symmetry therefore dictates that the corresponding components of $\chi$ must vanish, and prohibits photocurrent generation at normal incidence in systems described by the $C_\text{2}$ point groups or point group of higher symmetry. On the other hand, we clearly observe a photosignal on a fully strained $80$\,nm HgTe film with (001) orientation that exhibits two distinct peaks due to cyclotron resonance in the top and bottom surface layer of the film, see Fig.~\ref{fig4new}(d). Combined with the above symmetry argument, these observations suggest that in our sample there should be a mechanism that lowers the symmetry group down to $C_\text{s}$ or $C_1$. 

An obvious candidate for the mechanism responsible for the symmetry reduction is deformation caused by a lattice mismatch between different components of the grown heterostructure. However, in the case of (001)-grown structures such strain is not expected to lower the symmetry of interfaces. Indeed, the deformation of the HgTe film is defined by the CdTe layer. The in-plane lattice constant is the same throughout the structure and equal to $a_\text{CdTe}$, the lattice constant of CdTe. In this case the components of the strain tensor inside the HgTe film are given by $\epsilon_{xx}=\epsilon_{yy}=a_{\rm{CdTe}}/a_\text{\rm{HgTe}}-1$, $\epsilon_{xy}=\epsilon_{xz}=\epsilon_{yz}=0$, and $\epsilon_{zz}=-2 c_{12}(a_{\rm{CdTe}}/a_\text{\rm{HgTe}}-1)/c_{11}$, where $c_{11}$ and $c_{12}$ are elastic constants (here $x$, $y$ and $z$ denote the main crystallographic axes). This kind of deformation of the unit cell does not exclude any of the symmetry elements of the $C_\text{2v}$ point group. As illustrated in Fig.~\ref{fig8}(b), the two-fold rotation axis survives as an element of the symmetry group even under more general conditions which can in principle be superimposed by lattice matching in (001)-oriented structures for the strain tensor. In this case, $\epsilon_{xx}\neq \epsilon_{yy}$ and $\epsilon_{xy} \neq 0$, while other components $\epsilon_{xz}=0$, $\epsilon_{yz}=0$, and $\epsilon_{zz}=-c_{12}(\epsilon_{yy}+\epsilon_{xx})/c_{11}$ can be found from minimization of free elastic energy. As mentioned above, HgCdTe/HgTe/HgCdTe epilayers grown on GaAs substrate possess not only the relaxation stress induced by the mismatch of these materials but also significantly different thermal expansion coefficients of GaAs and epitaxial layers. However, by the same argument the related stress is not expected to lower the symmetry of the interface. We conclude that in the presence of strain photocurrents should still be forbidden under normal incidence of radiation due to high symmetry of the system. 
 		
Another possible reason for the symmetry reduction could be misorientation of the film surface from the (001)-plane by a small angle of several degrees.\cite{Ganichev2001} However, this contradicts the results of $X$-ray measurements showing that the deviations of both the substrate and the film from the (001)-plane do not exceed 5 angular minutes (not shown). While the origin of the symmetry reduction is presently unclear, the fact that the cyclotron masses of (001)-grown 80~nm HgTe films are $25\%$ lower  than the masses of the (013)-grown films may indicate that some structural defects are present in the films grown along [001] direction.

\section{Summary} 
\label{summary}

To summarize, our experiments demonstrate that topological states originating from the inverted band structure of HgTe persist in partially strained 200\,nm films grown on (013)-oriented GaAs substrate and behave similar to fully strained $70\div80$\,nm films with a bulk gap.\cite{Kozlov2014,Dantscher2015,Kozlov2016} This conclusion is supported by the observation of cyclotron resonances in transmission and photocurrent measurements. Experiments in tilted magnetic field prove that the resonances originate from two-dimensional surface states. Both methods indicate the presence of two cyclotron resonances positioned at slightly different magnetic fields. The cyclotron masses of approximately $0.03$ and $0.04\,m_0$ obtained from the positions of CRs correspond well to the values previously reported for the top and bottom surface states in 80\,nm HgTe films.\cite{Dantscher2015} This fact provides an evidence that topological surface states are robust to the existence of bulk carriers. Our findings are in agreement with the magnetotransport and capacitance data obtained on samples prepared from the same wafer and showing the existence of spin non-degenerate surface states.\cite{Savchenko2018} The remaining considerable stain in rather thick films grown on GaAs substrates, previously detected by $X$-ray diffraction,\cite{Savchenko2018} is discussed in our paper in terms of desirable extensions of the  Matthews model which may include the dislocation network formation as well as different thermal expansion coefficients of GaAs and HgCdTe/HgTe/CdHgTe/CdTe epitaxial layers. Our magnetotransport data show that the surface carriers of the system are high mobility electrons, coexisting with electrons and/or holes in the bulk. We further present a detailed theory of the cyclotron resonance assisted photogalvanic current in the surface states, and discuss specifics of the photocurrent generation in structures belonging to the $C_\text{1}$ point symmetry group.

\section{Acknowledgments}
\label{acknow} We thank S.A. Tarasenko and S.N. Danilov for helpful discussions. Support from the Deutsche Forschungsgemeinschaft (projects  SPP-1666 and DM1/4-1), the Elite Network of Bavaria (K-NW-2013-247) and the Volkswagen Stiftung Program is gratefully acknowledged. G.V.B. acknowledges support from the Russian Science Foundation (project no. 17-12-01265) and ''BASIS'' foundation. D.K. acknowledges support from the Russian Science Foundation (project no. 17-42-543336).

\newpage


\begin{thebibliography}{54}
\expandafter\ifx\csname natexlab\endcsname\relax\def\natexlab#1{#1}\fi
\expandafter\ifx\csname bibnamefont\endcsname\relax
  \def\bibnamefont#1{#1}\fi
\expandafter\ifx\csname bibfnamefont\endcsname\relax
  \def\bibfnamefont#1{#1}\fi
\expandafter\ifx\csname citenamefont\endcsname\relax
  \def\citenamefont#1{#1}\fi
\expandafter\ifx\csname url\endcsname\relax
  \def\url#1{\texttt{#1}}\fi
\expandafter\ifx\csname urlprefix\endcsname\relax\def\urlprefix{URL }\fi
\providecommand{\bibinfo}[2]{#2}
\providecommand{\eprint}[2][]{\url{#2}}

\bibitem[{\citenamefont{Hasan and Kane}(2010)}]{ColloquiumTI}
\bibinfo{author}{\bibfnamefont{M.~Z.} \bibnamefont{Hasan}} \bibnamefont{and}
  \bibinfo{author}{\bibfnamefont{C.~L.} \bibnamefont{Kane}},
  \bibinfo{journal}{Rev. Mod. Phys.} \textbf{\bibinfo{volume}{82}},
  \bibinfo{pages}{3045} (\bibinfo{year}{2010}).

\bibitem[{\citenamefont{Bernevig and Zhang}(2006)}]{Bernevig2006}
\bibinfo{author}{\bibfnamefont{B.~A.} \bibnamefont{Bernevig}} \bibnamefont{and}
  \bibinfo{author}{\bibfnamefont{S.~C.} \bibnamefont{Zhang}},
  \bibinfo{journal}{Phys. Rev. Lett.} \textbf{\bibinfo{volume}{96}},
  \bibinfo{pages}{106802} (\bibinfo{year}{2006}).

\bibitem[{\citenamefont{Br{\"{u}}ne et~al.}(2011)\citenamefont{Br{\"{u}}ne,
  Liu, Novik, Hankiewicz, Buhmann, Chen, Qi, Shen, Zhang, and
  Molenkamp}}]{Brune2011}
\bibinfo{author}{\bibfnamefont{C.}~\bibnamefont{Br{\"{u}}ne}},
  \bibinfo{author}{\bibfnamefont{C.~X.} \bibnamefont{Liu}},
  \bibinfo{author}{\bibfnamefont{E.~G.} \bibnamefont{Novik}},
  \bibinfo{author}{\bibfnamefont{E.~M.} \bibnamefont{Hankiewicz}},
  \bibinfo{author}{\bibfnamefont{H.}~\bibnamefont{Buhmann}},
  \bibinfo{author}{\bibfnamefont{Y.~L.} \bibnamefont{Chen}},
  \bibinfo{author}{\bibfnamefont{X.~L.} \bibnamefont{Qi}},
  \bibinfo{author}{\bibfnamefont{Z.~X.} \bibnamefont{Shen}},
  \bibinfo{author}{\bibfnamefont{S.~C.} \bibnamefont{Zhang}}, \bibnamefont{and}
  \bibinfo{author}{\bibfnamefont{L.~W.} \bibnamefont{Molenkamp}},
  \bibinfo{journal}{Phys. Rev. Lett.} \textbf{\bibinfo{volume}{106}},
  \bibinfo{pages}{126803} (\bibinfo{year}{2011}).

\bibitem[{\citenamefont{Kozlov et~al.}(2014)\citenamefont{Kozlov, Kvon,
  Olshanetsky, Mikhailov, Dvoretsky, and Weiss}}]{Kozlov2014}
\bibinfo{author}{\bibfnamefont{D.~A.} \bibnamefont{Kozlov}},
  \bibinfo{author}{\bibfnamefont{Z.~D.} \bibnamefont{Kvon}},
  \bibinfo{author}{\bibfnamefont{E.~B.} \bibnamefont{Olshanetsky}},
  \bibinfo{author}{\bibfnamefont{N.~N.} \bibnamefont{Mikhailov}},
  \bibinfo{author}{\bibfnamefont{S.~A.} \bibnamefont{Dvoretsky}},
  \bibnamefont{and} \bibinfo{author}{\bibfnamefont{D.}~\bibnamefont{Weiss}},
  \bibinfo{journal}{Phys. Rev. Lett.} \textbf{\bibinfo{volume}{112}},
  \bibinfo{pages}{196801} (\bibinfo{year}{2014}).

\bibitem[{\citenamefont{Wiedenmann et~al.}(2016)\citenamefont{Wiedenmann,
  Bocquillon, Deacon, Hartinger, Herrmann, Klapwijk, Maier, Ames, Br{\"{u}}ne,
  Gould et~al.}}]{Wiedenmann2016}
\bibinfo{author}{\bibfnamefont{J.}~\bibnamefont{Wiedenmann}},
  \bibinfo{author}{\bibfnamefont{E.}~\bibnamefont{Bocquillon}},
  \bibinfo{author}{\bibfnamefont{R.~S.} \bibnamefont{Deacon}},
  \bibinfo{author}{\bibfnamefont{S.}~\bibnamefont{Hartinger}},
  \bibinfo{author}{\bibfnamefont{O.}~\bibnamefont{Herrmann}},
  \bibinfo{author}{\bibfnamefont{T.~M.} \bibnamefont{Klapwijk}},
  \bibinfo{author}{\bibfnamefont{L.}~\bibnamefont{Maier}},
  \bibinfo{author}{\bibfnamefont{C.}~\bibnamefont{Ames}},
  \bibinfo{author}{\bibfnamefont{C.}~\bibnamefont{Br{\"{u}}ne}},
  \bibinfo{author}{\bibfnamefont{C.}~\bibnamefont{Gould}},
  \bibinfo{author}{\bibfnamefont{A.}~\bibnamefont{Oiwa}},
  \bibinfo{author}{\bibfnamefont{K.}~\bibnamefont{Ishibashi}},
  \bibinfo{author}{\bibfnamefont{S.}~\bibnamefont{Tarucha}},
  \bibinfo{author}{\bibfnamefont{H.}~\bibnamefont{Buhmann}}, \bibnamefont{and}
  \bibinfo{author}{\bibfnamefont{L.~W.} \bibnamefont{Molenkamp}},
  \bibinfo{journal}{Nature Commun.} \textbf{\bibinfo{volume}{7}},
  \bibinfo{pages}{10303} (\bibinfo{year}{2016}).

\bibitem[{\citenamefont{Wiedenmann et~al.}(2017)\citenamefont{Wiedenmann,
  Liebhaber, K{\"{u}}bert, Bocquillon, Burset, Ames, Buhmann, Klapwijk, and
  Molenkamp}}]{Wiedenmann2017}
\bibinfo{author}{\bibfnamefont{J.}~\bibnamefont{Wiedenmann}},
  \bibinfo{author}{\bibfnamefont{E.}~\bibnamefont{Liebhaber}},
  \bibinfo{author}{\bibfnamefont{J.}~\bibnamefont{K{\"{u}}bert}},
  \bibinfo{author}{\bibfnamefont{E.}~\bibnamefont{Bocquillon}},
  \bibinfo{author}{\bibfnamefont{P.}~\bibnamefont{Burset}},
  \bibinfo{author}{\bibfnamefont{C.}~\bibnamefont{Ames}},
  \bibinfo{author}{\bibfnamefont{H.}~\bibnamefont{Buhmann}},
  \bibinfo{author}{\bibfnamefont{T.~M.} \bibnamefont{Klapwijk}},
  \bibnamefont{and} \bibinfo{author}{\bibfnamefont{L.~W.}
  \bibnamefont{Molenkamp}}, \bibinfo{journal}{Phys. Rev. B}
  \textbf{\bibinfo{volume}{96}}, \bibinfo{pages}{165302}
  (\bibinfo{year}{2017}).

\bibitem[{\citenamefont{Maier et~al.}(2017)\citenamefont{Maier, Ziegler,
  Fischer, Kozlov, Kvon, Mikhailov, Dvoretsky, and Weiss}}]{Maier2017}
\bibinfo{author}{\bibfnamefont{H.}~\bibnamefont{Maier}},
  \bibinfo{author}{\bibfnamefont{J.}~\bibnamefont{Ziegler}},
  \bibinfo{author}{\bibfnamefont{R.}~\bibnamefont{Fischer}},
  \bibinfo{author}{\bibfnamefont{D.}~\bibnamefont{Kozlov}},
  \bibinfo{author}{\bibfnamefont{Z.~D.} \bibnamefont{Kvon}},
  \bibinfo{author}{\bibfnamefont{N.~N.} \bibnamefont{Mikhailov}},
  \bibinfo{author}{\bibfnamefont{S.~A.} \bibnamefont{Dvoretsky}},
  \bibnamefont{and} \bibinfo{author}{\bibfnamefont{D.}~\bibnamefont{Weiss}},
  \bibinfo{journal}{Nature Commun.} \textbf{\bibinfo{volume}{8}},
  \bibinfo{pages}{2023} (\bibinfo{year}{2017}).

\bibitem[{\citenamefont{Ziegler et~al.}(2018)\citenamefont{Ziegler, Kozlovsky,
  Gorini, Liu, Weish{\"{a}}upl, Maier, Fischer, Kozlov, Kvon, Mikhailov
  et~al.}}]{Ziegler2018}
\bibinfo{author}{\bibfnamefont{J.}~\bibnamefont{Ziegler}},
  \bibinfo{author}{\bibfnamefont{R.}~\bibnamefont{Kozlovsky}},
  \bibinfo{author}{\bibfnamefont{C.}~\bibnamefont{Gorini}},
  \bibinfo{author}{\bibfnamefont{M.-H.} \bibnamefont{Liu}},
  \bibinfo{author}{\bibfnamefont{S.}~\bibnamefont{Weish{\"{a}}upl}},
  \bibinfo{author}{\bibfnamefont{H.}~\bibnamefont{Maier}},
  \bibinfo{author}{\bibfnamefont{R.}~\bibnamefont{Fischer}},
  \bibinfo{author}{\bibfnamefont{D.~A.} \bibnamefont{Kozlov}},
  \bibinfo{author}{\bibfnamefont{Z.~D.} \bibnamefont{Kvon}},
  \bibinfo{author}{\bibfnamefont{N.~N.} \bibnamefont{Mikhailov}},
  \bibinfo{author}{\bibfnamefont{S.~A.} \bibnamefont{Dvoretsky}},
  \bibinfo{author}{\bibfnamefont{K.}~\bibnamefont{Richter}}, \bibnamefont{and}
  \bibinfo{author}{\bibfnamefont{D.}~\bibnamefont{Weiss}},
  \bibinfo{journal}{Phys. Rev. B} \textbf{\bibinfo{volume}{97}},
  \bibinfo{pages}{035157} (\bibinfo{year}{2018}).

\bibitem[{\citenamefont{Kozlov et~al.}(2016)\citenamefont{Kozlov, Bauer,
  Ziegler, Fischer, Savchenko, Kvon, Mikhailov, Dvoretsky, and
  Weiss}}]{Kozlov2016}
\bibinfo{author}{\bibfnamefont{D.~A.} \bibnamefont{Kozlov}},
  \bibinfo{author}{\bibfnamefont{D.}~\bibnamefont{Bauer}},
  \bibinfo{author}{\bibfnamefont{J.}~\bibnamefont{Ziegler}},
  \bibinfo{author}{\bibfnamefont{R.}~\bibnamefont{Fischer}},
  \bibinfo{author}{\bibfnamefont{M.~L.} \bibnamefont{Savchenko}},
  \bibinfo{author}{\bibfnamefont{Z.~D.} \bibnamefont{Kvon}},
  \bibinfo{author}{\bibfnamefont{N.~N.} \bibnamefont{Mikhailov}},
  \bibinfo{author}{\bibfnamefont{S.~A.} \bibnamefont{Dvoretsky}},
  \bibnamefont{and} \bibinfo{author}{\bibfnamefont{D.}~\bibnamefont{Weiss}},
  \bibinfo{journal}{Phys. Rev. Lett.} \textbf{\bibinfo{volume}{116}},
  \bibinfo{pages}{166802} (\bibinfo{year}{2016}).

\bibitem[{\citenamefont{Shuvaev et~al.}(2012)\citenamefont{Shuvaev, Astakhov,
  Br{\"{u}}ne, Buhmann, Molenkamp, and Pimenov}}]{Shuvaev2012}
\bibinfo{author}{\bibfnamefont{A.}~\bibnamefont{Shuvaev}},
  \bibinfo{author}{\bibfnamefont{G.}~\bibnamefont{Astakhov}},
  \bibinfo{author}{\bibfnamefont{C.}~\bibnamefont{Br{\"{u}}ne}},
  \bibinfo{author}{\bibfnamefont{H.}~\bibnamefont{Buhmann}},
  \bibinfo{author}{\bibfnamefont{L.~W.} \bibnamefont{Molenkamp}},
  \bibnamefont{and} \bibinfo{author}{\bibfnamefont{A.}~\bibnamefont{Pimenov}},
  \bibinfo{journal}{Semicond. Sci. Technol.} \textbf{\bibinfo{volume}{27}},
  \bibinfo{pages}{124004} (\bibinfo{year}{2012}).

\bibitem[{\citenamefont{Shuvaev et~al.}(2013)\citenamefont{Shuvaev, Pimenov,
  Astakhov, M{\"{u}}hlbauer, Br{\"{u}}ne, Buhmann, and
  Molenkamp}}]{Shuvaev2013a}
\bibinfo{author}{\bibfnamefont{A.}~\bibnamefont{Shuvaev}},
  \bibinfo{author}{\bibfnamefont{A.}~\bibnamefont{Pimenov}},
  \bibinfo{author}{\bibfnamefont{G.~V.} \bibnamefont{Astakhov}},
  \bibinfo{author}{\bibfnamefont{M.}~\bibnamefont{M{\"{u}}hlbauer}},
  \bibinfo{author}{\bibfnamefont{C.}~\bibnamefont{Br{\"{u}}ne}},
  \bibinfo{author}{\bibfnamefont{H.}~\bibnamefont{Buhmann}}, \bibnamefont{and}
  \bibinfo{author}{\bibfnamefont{L.~W.} \bibnamefont{Molenkamp}},
  \bibinfo{journal}{Appl. Phys. Lett.} \textbf{\bibinfo{volume}{102}},
  \bibinfo{pages}{241902} (\bibinfo{year}{2013}).

\bibitem[{\citenamefont{Dantscher et~al.}(2015)\citenamefont{Dantscher, Kozlov,
  Olbrich, Zoth, Faltermeier, Lindner, Budkin, Tarasenko, Bel'kov, Kvon
  et~al.}}]{Dantscher2015}
\bibinfo{author}{\bibfnamefont{K.~M.} \bibnamefont{Dantscher}},
  \bibinfo{author}{\bibfnamefont{D.~A.} \bibnamefont{Kozlov}},
  \bibinfo{author}{\bibfnamefont{P.}~\bibnamefont{Olbrich}},
  \bibinfo{author}{\bibfnamefont{C.}~\bibnamefont{Zoth}},
  \bibinfo{author}{\bibfnamefont{P.}~\bibnamefont{Faltermeier}},
  \bibinfo{author}{\bibfnamefont{M.}~\bibnamefont{Lindner}},
  \bibinfo{author}{\bibfnamefont{G.~V.} \bibnamefont{Budkin}},
  \bibinfo{author}{\bibfnamefont{S.~A.} \bibnamefont{Tarasenko}},
  \bibinfo{author}{\bibfnamefont{V.~V.} \bibnamefont{Bel'kov}},
  \bibinfo{author}{\bibfnamefont{Z.~D.} \bibnamefont{Kvon}},
  \bibinfo{author}{\bibfnamefont{N.~N.} \bibnamefont{Mikhailov}},
  \bibinfo{author}{\bibfnamefont{S.~A.} \bibnamefont{Dvoretsky}},
  \bibinfo{author}{\bibfnamefont{D.}~\bibnamefont{Weiss}},
  \bibinfo{author}{\bibfnamefont{B.}~\bibnamefont{Jenichen}}, \bibnamefont{and}
  \bibinfo{author}{\bibfnamefont{S.~D.} \bibnamefont{Ganichev}},
  \bibinfo{journal}{Phys. Rev. B} \textbf{\bibinfo{volume}{92}},
  \bibinfo{pages}{165314} (\bibinfo{year}{2015}).

\bibitem[{\citenamefont{Wu et~al.}(2016)\citenamefont{Wu, Salehi, Koirala,
  Moon, Oh, and Armitage}}]{Wu2016}
\bibinfo{author}{\bibfnamefont{L.}~\bibnamefont{Wu}},
  \bibinfo{author}{\bibfnamefont{M.}~\bibnamefont{Salehi}},
  \bibinfo{author}{\bibfnamefont{N.}~\bibnamefont{Koirala}},
  \bibinfo{author}{\bibfnamefont{J.}~\bibnamefont{Moon}},
  \bibinfo{author}{\bibfnamefont{S.}~\bibnamefont{Oh}}, \bibnamefont{and}
  \bibinfo{author}{\bibfnamefont{N.~P.} \bibnamefont{Armitage}},
  \bibinfo{journal}{Science} \textbf{\bibinfo{volume}{354}},
  \bibinfo{pages}{1124} (\bibinfo{year}{2016}).

\bibitem[{\citenamefont{Savchenko et~al.}(2018)\citenamefont{Savchenko, Kozlov,
  Vasilev, Kvon, Mikhailov, Dvoretsky, and Kolesnikov}}]{Savchenko2018}
\bibinfo{author}{\bibfnamefont{M.~L.} \bibnamefont{Savchenko}},
  \bibinfo{author}{\bibfnamefont{D.~A.} \bibnamefont{Kozlov}},
  \bibinfo{author}{\bibfnamefont{N.~N.} \bibnamefont{Vasilev}},
  \bibinfo{author}{\bibfnamefont{Z.~D.} \bibnamefont{Kvon}},
  \bibinfo{author}{\bibfnamefont{N.~N.} \bibnamefont{Mikhailov}},
  \bibinfo{author}{\bibfnamefont{S.~A.} \bibnamefont{Dvoretsky}},
  \bibnamefont{and} \bibinfo{author}{\bibfnamefont{A.~V.}
  \bibnamefont{Kolesnikov}}, \bibinfo{journal}{arXiv:1811.03385}
  (\bibinfo{year}{2018}).

\bibitem[{\citenamefont{Kvon et~al.}(2008)\citenamefont{Kvon, Olshanetsky,
  Kozlov, Mikhailov, and Dvoretskii}}]{Kvon2008}
\bibinfo{author}{\bibfnamefont{Z.~D.} \bibnamefont{Kvon}},
  \bibinfo{author}{\bibfnamefont{E.~B.} \bibnamefont{Olshanetsky}},
  \bibinfo{author}{\bibfnamefont{D.~A.} \bibnamefont{Kozlov}},
  \bibinfo{author}{\bibfnamefont{N.~N.} \bibnamefont{Mikhailov}},
  \bibnamefont{and} \bibinfo{author}{\bibfnamefont{S.~A.}
  \bibnamefont{Dvoretskii}}, \bibinfo{journal}{{JETP} Lett.}
  \textbf{\bibinfo{volume}{87}}, \bibinfo{pages}{502} (\bibinfo{year}{2008}).

\bibitem[{\citenamefont{Ganichev et~al.}(2009)\citenamefont{Ganichev,
  Tarasenko, Bel'kov, Olbrich, Eder, Yakovlev, Kolkovsky, Zaleszczyk,
  Karczewski, Wojtowicz et~al.}}]{Ganichev2009}
\bibinfo{author}{\bibfnamefont{S.}~\bibnamefont{Ganichev}},
  \bibinfo{author}{\bibfnamefont{S.}~\bibnamefont{Tarasenko}},
  \bibinfo{author}{\bibfnamefont{V.}~\bibnamefont{Bel'kov}},
  \bibinfo{author}{\bibfnamefont{P.}~\bibnamefont{Olbrich}},
  \bibinfo{author}{\bibfnamefont{W.}~\bibnamefont{Eder}},
  \bibinfo{author}{\bibfnamefont{D.}~\bibnamefont{Yakovlev}},
  \bibinfo{author}{\bibfnamefont{V.}~\bibnamefont{Kolkovsky}},
  \bibinfo{author}{\bibfnamefont{W.}~\bibnamefont{Zaleszczyk}},
  \bibinfo{author}{\bibfnamefont{G.}~\bibnamefont{Karczewski}},
  \bibinfo{author}{\bibfnamefont{T.}~\bibnamefont{Wojtowicz}},
  \bibnamefont{and} \bibinfo{author}{\bibfnamefont{D.}~\bibnamefont{Weiss}},
  \bibinfo{journal}{Phys. Rev. Lett.} \textbf{\bibinfo{volume}{102}},
  \bibinfo{pages}{156602} (\bibinfo{year}{2009}).

\bibitem[{\citenamefont{Olbrich et~al.}(2011)\citenamefont{Olbrich, Karch,
  Ivchenko, Kamann, März, Fehrenbacher, Weiss, and Ganichev}}]{Olbrich2011}
\bibinfo{author}{\bibfnamefont{P.}~\bibnamefont{Olbrich}},
  \bibinfo{author}{\bibfnamefont{J.}~\bibnamefont{Karch}},
  \bibinfo{author}{\bibfnamefont{E.~L.} \bibnamefont{Ivchenko}},
  \bibinfo{author}{\bibfnamefont{J.}~\bibnamefont{Kamann}},
  \bibinfo{author}{\bibfnamefont{B.}~\bibnamefont{März}},
  \bibinfo{author}{\bibfnamefont{M.}~\bibnamefont{Fehrenbacher}},
  \bibinfo{author}{\bibfnamefont{D.}~\bibnamefont{Weiss}}, \bibnamefont{and}
  \bibinfo{author}{\bibfnamefont{S.~D.} \bibnamefont{Ganichev}},
  \bibinfo{journal}{Phys. Rev. B} \textbf{\bibinfo{volume}{83}},
  \bibinfo{pages}{235439} (\bibinfo{year}{2011}).

\bibitem[{\citenamefont{Olbrich et~al.}(2014)\citenamefont{Olbrich, Golub,
  Herrmann, Danilov, Plank, Bel'kov, Mussler, Weyrich, Schneider, Kampmeier
  et~al.}}]{Olbrich2014}
\bibinfo{author}{\bibfnamefont{P.}~\bibnamefont{Olbrich}},
  \bibinfo{author}{\bibfnamefont{L.~E.} \bibnamefont{Golub}},
  \bibinfo{author}{\bibfnamefont{T.}~\bibnamefont{Herrmann}},
  \bibinfo{author}{\bibfnamefont{S.~N.} \bibnamefont{Danilov}},
  \bibinfo{author}{\bibfnamefont{H.}~\bibnamefont{Plank}},
  \bibinfo{author}{\bibfnamefont{V.~V.} \bibnamefont{Bel'kov}},
  \bibinfo{author}{\bibfnamefont{G.}~\bibnamefont{Mussler}},
  \bibinfo{author}{\bibfnamefont{C.}~\bibnamefont{Weyrich}},
  \bibinfo{author}{\bibfnamefont{C.~M.} \bibnamefont{Schneider}},
  \bibinfo{author}{\bibfnamefont{J.}~\bibnamefont{Kampmeier}},
  \bibinfo{author}{\bibfnamefont{D.}~\bibnamefont{Gr\"utzmacher}},
  \bibinfo{author}{\bibfnamefont{L.}~\bibnamefont{Plucinski}},
  \bibinfo{author}{\bibfnamefont{M.}~\bibnamefont{Eschbach}}, \bibnamefont{and}
  \bibinfo{author}{\bibfnamefont{S.~D.} \bibnamefont{Ganichev}},
  \bibinfo{journal}{Phys. Rev. Lett.} \textbf{\bibinfo{volume}{113}},
  \bibinfo{pages}{096601} (\bibinfo{year}{2014}).

\bibitem[{\citenamefont{Ganichev et~al.}(1982)\citenamefont{Ganichev,
  Emel'yanov, and Yaroshetskii}}]{Ganichev1982}
\bibinfo{author}{\bibfnamefont{S.}~\bibnamefont{Ganichev}},
  \bibinfo{author}{\bibfnamefont{S.}~\bibnamefont{Emel'yanov}},
  \bibnamefont{and}
  \bibinfo{author}{\bibfnamefont{I.}~\bibnamefont{Yaroshetskii}},
  \bibinfo{journal}{JETP Lett.} \textbf{\bibinfo{volume}{35}},
  \bibinfo{pages}{368} (\bibinfo{year}{1982}).

\bibitem[{\citenamefont{Ganichev}(1999)}]{Ganichev1999}
\bibinfo{author}{\bibfnamefont{S.}~\bibnamefont{Ganichev}},
  \bibinfo{journal}{Physica B} \textbf{\bibinfo{volume}{273-274}},
  \bibinfo{pages}{737} (\bibinfo{year}{1999}).

\bibitem[{\citenamefont{Ganichev and Prettl}(2005)}]{Ganichev2005}
\bibinfo{author}{\bibfnamefont{S.}~\bibnamefont{Ganichev}} \bibnamefont{and}
  \bibinfo{author}{\bibfnamefont{W.}~\bibnamefont{Prettl}},
  \emph{\bibinfo{title}{Intense Terahertz Excitation of Semiconductors}}
  (\bibinfo{publisher}{Oxford University Press}, \bibinfo{year}{2005}).

\bibitem[{foo()}]{footnote1}
\bibinfo{note}{A small deviation from monotonic behavior of the photocurrent,
  which is seen for CR passive magnetic field polarity (magnitude of peaks is
  about 5 times smaller than that for the CR active polarity) is attributed to
  non-perfect quarter plate in our disposal. The magnitude of the resonant part
  of the signals (subtracting the monotonic non-resonant photocurrent) excited
  by circularly polarized radiation is approximately twice larger than in the
  case of linearly polarized radiation of the same intensity, see
  Fig.~\ref{fig3}(a).}

\bibitem[{\citenamefont{Wittmann et~al.}(2010)\citenamefont{Wittmann, Danilov,
  Bel'kov, Tarasenko, Novik, Buhmann, Br\"une, Molenkamp, Kvon, Mikhailov
  et~al.}}]{Wittmann2010}
\bibinfo{author}{\bibfnamefont{B.}~\bibnamefont{Wittmann}},
  \bibinfo{author}{\bibfnamefont{S.}~\bibnamefont{Danilov}},
  \bibinfo{author}{\bibfnamefont{V.}~\bibnamefont{Bel'kov}},
  \bibinfo{author}{\bibfnamefont{S.}~\bibnamefont{Tarasenko}},
  \bibinfo{author}{\bibfnamefont{E.}~\bibnamefont{Novik}},
  \bibinfo{author}{\bibfnamefont{H.}~\bibnamefont{Buhmann}},
  \bibinfo{author}{\bibfnamefont{C.}~\bibnamefont{Br\"une}},
  \bibinfo{author}{\bibfnamefont{L.}~\bibnamefont{Molenkamp}},
  \bibinfo{author}{\bibfnamefont{Z.}~\bibnamefont{Kvon}},
  \bibinfo{author}{\bibfnamefont{N.}~\bibnamefont{Mikhailov}},
  \bibinfo{author}{\bibfnamefont{S.}~\bibnamefont{Dvoretsky}},
  \bibinfo{author}{\bibfnamefont{N.}~\bibnamefont{Vinh}},
  \bibinfo{author}{\bibfnamefont{A.}~\bibnamefont{van~der Meer}},
  \bibinfo{author}{\bibfnamefont{B.}~\bibnamefont{Murdin}}, \bibnamefont{and}
  \bibinfo{author}{\bibfnamefont{S.}~\bibnamefont{Ganichev}},
  \bibinfo{journal}{Semicond. Sci. Technol.} \textbf{\bibinfo{volume}{25}},
  \bibinfo{pages}{095005} (\bibinfo{year}{2010}).

\bibitem[{\citenamefont{Matthews}(1975)}]{Matthews1975}
\bibinfo{author}{\bibfnamefont{J.~W.} \bibnamefont{Matthews}},
  \bibinfo{journal}{J. Vac. Sci. Technol.} \textbf{\bibinfo{volume}{12}},
  \bibinfo{pages}{126} (\bibinfo{year}{1975}).

\bibitem[{\citenamefont{Pichaud et~al.}(2002)\citenamefont{Pichaud, Burle,
  Putero-Vuaroqueaux, and Curtil}}]{Pichaud2002}
\bibinfo{author}{\bibfnamefont{B.}~\bibnamefont{Pichaud}},
  \bibinfo{author}{\bibfnamefont{N.}~\bibnamefont{Burle}},
  \bibinfo{author}{\bibfnamefont{M.}~\bibnamefont{Putero-Vuaroqueaux}},
  \bibnamefont{and} \bibinfo{author}{\bibfnamefont{C.}~\bibnamefont{Curtil}},
  \bibinfo{journal}{J. Phys.: Condens. Matter} \textbf{\bibinfo{volume}{14}},
  \bibinfo{pages}{13255} (\bibinfo{year}{2002}).

\bibitem[{\citenamefont{France et~al.}(2011)\citenamefont{France, Jiang, and
  Ptak}}]{France2011a}
\bibinfo{author}{\bibfnamefont{R.}~\bibnamefont{France}},
  \bibinfo{author}{\bibfnamefont{C.-S.} \bibnamefont{Jiang}}, \bibnamefont{and}
  \bibinfo{author}{\bibfnamefont{A.~J.} \bibnamefont{Ptak}},
  \bibinfo{journal}{Appl. Phys. Lett.} \textbf{\bibinfo{volume}{98}},
  \bibinfo{pages}{101908} (\bibinfo{year}{2011}).

\bibitem[{\citenamefont{France and Ptak}(2011)}]{France2011}
\bibinfo{author}{\bibfnamefont{R.}~\bibnamefont{France}} \bibnamefont{and}
  \bibinfo{author}{\bibfnamefont{A.~J.} \bibnamefont{Ptak}},
  \bibinfo{journal}{J. Vac. Sci. Technol. B} \textbf{\bibinfo{volume}{29}},
  \bibinfo{pages}{03C115} (\bibinfo{year}{2011}).

\bibitem[{\citenamefont{Bolkhovityanov
  et~al.}(2001)\citenamefont{Bolkhovityanov, Pchelyakov, and
  Chikichev}}]{Bolkhovityanov2001}
\bibinfo{author}{\bibfnamefont{Y.}~\bibnamefont{Bolkhovityanov}},
  \bibinfo{author}{\bibfnamefont{O.~P.} \bibnamefont{Pchelyakov}},
  \bibnamefont{and}
  \bibinfo{author}{\bibfnamefont{S.}~\bibnamefont{Chikichev}},
  \bibinfo{journal}{Physics-Uspekhi} \textbf{\bibinfo{volume}{171}},
  \bibinfo{pages}{689} (\bibinfo{year}{2001}).

\bibitem[{\citenamefont{Basson and Booyens}(1983)}]{Basson1983}
\bibinfo{author}{\bibfnamefont{J.~H.} \bibnamefont{Basson}} \bibnamefont{and}
  \bibinfo{author}{\bibfnamefont{H.}~\bibnamefont{Booyens}},
  \bibinfo{journal}{Phys. Status Solidi (a)} \textbf{\bibinfo{volume}{80}},
  \bibinfo{pages}{663} (\bibinfo{year}{1983}).

\bibitem[{\citenamefont{Ballet et~al.}(2014)\citenamefont{Ballet, Thomas,
  Baudry, Bouvier, Crauste, Meunier, Badano, Veillerot, Barnes, Jouneau
  et~al.}}]{Ballet2014}
\bibinfo{author}{\bibfnamefont{P.}~\bibnamefont{Ballet}},
  \bibinfo{author}{\bibfnamefont{C.}~\bibnamefont{Thomas}},
  \bibinfo{author}{\bibfnamefont{X.}~\bibnamefont{Baudry}},
  \bibinfo{author}{\bibfnamefont{C.}~\bibnamefont{Bouvier}},
  \bibinfo{author}{\bibfnamefont{O.}~\bibnamefont{Crauste}},
  \bibinfo{author}{\bibfnamefont{T.}~\bibnamefont{Meunier}},
  \bibinfo{author}{\bibfnamefont{G.}~\bibnamefont{Badano}},
  \bibinfo{author}{\bibfnamefont{M.}~\bibnamefont{Veillerot}},
  \bibinfo{author}{\bibfnamefont{J.~P.} \bibnamefont{Barnes}},
  \bibinfo{author}{\bibfnamefont{P.~H.} \bibnamefont{Jouneau}},
  \bibnamefont{and} \bibinfo{author}{\bibfnamefont{L.~P.} \bibnamefont{Levy}},
  \bibinfo{journal}{J. Electr. Mat.} \textbf{\bibinfo{volume}{43}},
  \bibinfo{pages}{2955} (\bibinfo{year}{2014}).

\bibitem[{\citenamefont{Berding et~al.}(2000)\citenamefont{Berding, Nix,
  Rhiger, Sen, and Sher}}]{Berding2000}
\bibinfo{author}{\bibfnamefont{M.~A.} \bibnamefont{Berding}},
  \bibinfo{author}{\bibfnamefont{W.~D.} \bibnamefont{Nix}},
  \bibinfo{author}{\bibfnamefont{D.~R.} \bibnamefont{Rhiger}},
  \bibinfo{author}{\bibfnamefont{S.}~\bibnamefont{Sen}}, \bibnamefont{and}
  \bibinfo{author}{\bibfnamefont{A.}~\bibnamefont{Sher}}, \bibinfo{journal}{J.
  Electr. Mat.} \textbf{\bibinfo{volume}{29}}, \bibinfo{pages}{676}
  (\bibinfo{year}{2000}).

\bibitem[{\citenamefont{Sidorov et~al.}(2015)\citenamefont{Sidorov, Yakushev,
  Varavin, Kolesnikov, Trukhanov, Sabinina, and Loshkarev}}]{Sidorov2015}
\bibinfo{author}{\bibfnamefont{Y.~G.} \bibnamefont{Sidorov}},
  \bibinfo{author}{\bibfnamefont{M.~V.} \bibnamefont{Yakushev}},
  \bibinfo{author}{\bibfnamefont{V.~S.} \bibnamefont{Varavin}},
  \bibinfo{author}{\bibfnamefont{A.~V.} \bibnamefont{Kolesnikov}},
  \bibinfo{author}{\bibfnamefont{E.~M.} \bibnamefont{Trukhanov}},
  \bibinfo{author}{\bibfnamefont{I.~V.} \bibnamefont{Sabinina}},
  \bibnamefont{and} \bibinfo{author}{\bibfnamefont{I.~D.}
  \bibnamefont{Loshkarev}}, \bibinfo{journal}{Phys. Solid State}
  \textbf{\bibinfo{volume}{57}}, \bibinfo{pages}{2151} (\bibinfo{year}{2015}).

\bibitem[{\citenamefont{Herrmann et~al.}(2016)\citenamefont{Herrmann, Dmitriev,
  Kozlov, Schneider, Jentzsch, Kvon, Olbrich, Bel{\textquotesingle}kov, Bayer,
  Schuh et~al.}}]{Herrmann2016}
\bibinfo{author}{\bibfnamefont{T.}~\bibnamefont{Herrmann}},
  \bibinfo{author}{\bibfnamefont{I.~A.} \bibnamefont{Dmitriev}},
  \bibinfo{author}{\bibfnamefont{D.~A.} \bibnamefont{Kozlov}},
  \bibinfo{author}{\bibfnamefont{M.}~\bibnamefont{Schneider}},
  \bibinfo{author}{\bibfnamefont{B.}~\bibnamefont{Jentzsch}},
  \bibinfo{author}{\bibfnamefont{Z.~D.} \bibnamefont{Kvon}},
  \bibinfo{author}{\bibfnamefont{P.}~\bibnamefont{Olbrich}},
  \bibinfo{author}{\bibfnamefont{V.~V.}
  \bibnamefont{Bel{\textquotesingle}kov}},
  \bibinfo{author}{\bibfnamefont{A.}~\bibnamefont{Bayer}},
  \bibinfo{author}{\bibfnamefont{D.}~\bibnamefont{Schuh}},
  \bibinfo{author}{\bibfnamefont{D.}~\bibnamefont{Bougeard}},
  \bibinfo{author}{\bibfnamefont{T.}~\bibnamefont{Kuczmik}},
  \bibinfo{author}{\bibfnamefont{M.}~\bibnamefont{Oltscher}},
  \bibinfo{author}{\bibfnamefont{D.}~\bibnamefont{Weiss}}, \bibnamefont{and}
  \bibinfo{author}{\bibfnamefont{S.~D.} \bibnamefont{Ganichev}},
  \bibinfo{journal}{Phys. Rev. B} \textbf{\bibinfo{volume}{94}},
  \bibinfo{pages}{081301(R)} (\bibinfo{year}{2016}).

\bibitem[{\citenamefont{Abstreiter et~al.}(1976)\citenamefont{Abstreiter,
  Kotthaus, Koch, and Dorda}}]{Abstreiter1976}
\bibinfo{author}{\bibfnamefont{G.}~\bibnamefont{Abstreiter}},
  \bibinfo{author}{\bibfnamefont{J.~P.} \bibnamefont{Kotthaus}},
  \bibinfo{author}{\bibfnamefont{J.~F.} \bibnamefont{Koch}}, \bibnamefont{and}
  \bibinfo{author}{\bibfnamefont{G.}~\bibnamefont{Dorda}},
  \bibinfo{journal}{Phys. Rev. B} \textbf{\bibinfo{volume}{14}},
  \bibinfo{pages}{2480} (\bibinfo{year}{1976}).

\bibitem[{\citenamefont{{Chiu} et~al.}(1976)\citenamefont{{Chiu}, {Lee}, and
  {Quinn}}}]{Chiu1976}
\bibinfo{author}{\bibfnamefont{K.~W.} \bibnamefont{{Chiu}}},
  \bibinfo{author}{\bibfnamefont{T.~K.} \bibnamefont{{Lee}}}, \bibnamefont{and}
  \bibinfo{author}{\bibfnamefont{J.~J.} \bibnamefont{{Quinn}}},
  \bibinfo{journal}{Surf. Sci.} \textbf{\bibinfo{volume}{58}},
  \bibinfo{pages}{182} (\bibinfo{year}{1976}).

\bibitem[{\citenamefont{Mikhailov}(2004)}]{Mikhailov2004}
\bibinfo{author}{\bibfnamefont{S.~A.} \bibnamefont{Mikhailov}},
  \bibinfo{journal}{Phys. Rev. B} \textbf{\bibinfo{volume}{70}},
  \bibinfo{pages}{165311} (\bibinfo{year}{2004}).

\bibitem[{\citenamefont{Zhang et~al.}(2014)\citenamefont{Zhang, Arikawa, Kato,
  Reno, Pan, Watson, Manfra, Zudov, Tokman, Erukhimova et~al.}}]{Zhang2014}
\bibinfo{author}{\bibfnamefont{Q.}~\bibnamefont{Zhang}},
  \bibinfo{author}{\bibfnamefont{T.}~\bibnamefont{Arikawa}},
  \bibinfo{author}{\bibfnamefont{E.}~\bibnamefont{Kato}},
  \bibinfo{author}{\bibfnamefont{J.~L.} \bibnamefont{Reno}},
  \bibinfo{author}{\bibfnamefont{W.}~\bibnamefont{Pan}},
  \bibinfo{author}{\bibfnamefont{J.~D.} \bibnamefont{Watson}},
  \bibinfo{author}{\bibfnamefont{M.~J.} \bibnamefont{Manfra}},
  \bibinfo{author}{\bibfnamefont{M.~A.} \bibnamefont{Zudov}},
  \bibinfo{author}{\bibfnamefont{M.}~\bibnamefont{Tokman}},
  \bibinfo{author}{\bibfnamefont{M.}~\bibnamefont{Erukhimova}},
  \bibinfo{author}{\bibfnamefont{A.}~\bibnamefont{Belyanin}}, \bibnamefont{and}
  \bibinfo{author}{\bibfnamefont{J.}~\bibnamefont{Kono}},
  \bibinfo{journal}{Phys. Rev. Lett.} \textbf{\bibinfo{volume}{113}},
  \bibinfo{pages}{047601} (\bibinfo{year}{2014}).

\bibitem[{\citenamefont{Stachel et~al.}(2014)\citenamefont{Stachel, Budkin,
  Hagner, Bel'kov, Glazov, Tarasenko, Clowes, Ashley, Gilbertson, and
  Ganichev}}]{Stachel2014}
\bibinfo{author}{\bibfnamefont{S.}~\bibnamefont{Stachel}},
  \bibinfo{author}{\bibfnamefont{G.~V.} \bibnamefont{Budkin}},
  \bibinfo{author}{\bibfnamefont{U.}~\bibnamefont{Hagner}},
  \bibinfo{author}{\bibfnamefont{V.}~\bibnamefont{Bel'kov}},
  \bibinfo{author}{\bibfnamefont{M.}~\bibnamefont{Glazov}},
  \bibinfo{author}{\bibfnamefont{S.}~\bibnamefont{Tarasenko}},
  \bibinfo{author}{\bibfnamefont{S.~K.} \bibnamefont{Clowes}},
  \bibinfo{author}{\bibfnamefont{T.}~\bibnamefont{Ashley}},
  \bibinfo{author}{\bibfnamefont{A.~M.} \bibnamefont{Gilbertson}},
  \bibnamefont{and} \bibinfo{author}{\bibfnamefont{S.}~\bibnamefont{Ganichev}},
  \bibinfo{journal}{Phys. Rev. B} \textbf{\bibinfo{volume}{89}},
  \bibinfo{pages}{115435} (\bibinfo{year}{2014}).

\bibitem[{\citenamefont{Olbrich et~al.}(2013)\citenamefont{Olbrich, Zoth,
  Vierling, Dantscher, Budkin, Tarasenko, Bel'kov, Kozlov, Kvon, Mikhailov
  et~al.}}]{Olbrich2013}
\bibinfo{author}{\bibfnamefont{P.}~\bibnamefont{Olbrich}},
  \bibinfo{author}{\bibfnamefont{C.}~\bibnamefont{Zoth}},
  \bibinfo{author}{\bibfnamefont{P.}~\bibnamefont{Vierling}},
  \bibinfo{author}{\bibfnamefont{K.-M.} \bibnamefont{Dantscher}},
  \bibinfo{author}{\bibfnamefont{G.~V.} \bibnamefont{Budkin}},
  \bibinfo{author}{\bibfnamefont{S.~A.} \bibnamefont{Tarasenko}},
  \bibinfo{author}{\bibfnamefont{V.~V.} \bibnamefont{Bel'kov}},
  \bibinfo{author}{\bibfnamefont{D.~A.} \bibnamefont{Kozlov}},
  \bibinfo{author}{\bibfnamefont{Z.~D.} \bibnamefont{Kvon}},
  \bibinfo{author}{\bibfnamefont{N.~N.} \bibnamefont{Mikhailov}},
  \bibinfo{author}{\bibfnamefont{S.~A.} \bibnamefont{Dvoretsky}},
  \bibnamefont{and} \bibinfo{author}{\bibfnamefont{S.~D.}
  \bibnamefont{Ganichev}}, \bibinfo{journal}{Phys. Rev. B}
  \textbf{\bibinfo{volume}{87}}, \bibinfo{pages}{235439}
  (\bibinfo{year}{2013}).

\bibitem[{\citenamefont{Budkin and Tarasenko}(2016)}]{budkin2016}
\bibinfo{author}{\bibfnamefont{G.~V.} \bibnamefont{Budkin}} \bibnamefont{and}
  \bibinfo{author}{\bibfnamefont{S.~A.} \bibnamefont{Tarasenko}},
  \bibinfo{journal}{Phys. Rev. B} \textbf{\bibinfo{volume}{93}},
  \bibinfo{pages}{075306} (\bibinfo{year}{2016}).

\bibitem[{\citenamefont{Budkin et~al.}(2015)\citenamefont{Budkin, Tarasenko,
  Bel'kov, Dantscher, Kozlov, Olbrich, Zoth, Faltermeier, Lindner, Kvon
  et~al.}}]{budkinnano2015}
\bibinfo{author}{\bibfnamefont{G.~V.} \bibnamefont{Budkin}},
  \bibinfo{author}{\bibfnamefont{S.~A.} \bibnamefont{Tarasenko}},
  \bibinfo{author}{\bibfnamefont{V.~V.} \bibnamefont{Bel'kov}},
  \bibinfo{author}{\bibfnamefont{K.-M.} \bibnamefont{Dantscher}},
  \bibinfo{author}{\bibfnamefont{D.~A.} \bibnamefont{Kozlov}},
  \bibinfo{author}{\bibfnamefont{P.}~\bibnamefont{Olbrich}},
  \bibinfo{author}{\bibfnamefont{C.}~\bibnamefont{Zoth}},
  \bibinfo{author}{\bibfnamefont{P.}~\bibnamefont{Faltermeier}},
  \bibinfo{author}{\bibfnamefont{M.}~\bibnamefont{Lindner}},
  \bibinfo{author}{\bibfnamefont{Z.~D.} \bibnamefont{Kvon}},
  \bibinfo{author}{\bibfnamefont{N.~N.} \bibnamefont{Mikhailov}},
  \bibinfo{author}{\bibfnamefont{S.~A.} \bibnamefont{Dvoretsky}},
  \bibinfo{author}{\bibfnamefont{D.}~\bibnamefont{Weiss}}, \bibnamefont{and}
  \bibinfo{author}{\bibfnamefont{S.~D.} \bibnamefont{Ganichev}}, in
  \emph{\bibinfo{booktitle}{23rd Int. Symp. "Nanostructures: Physics and
  Technology"}} (\bibinfo{publisher}{Academic University Publishing},
  \bibinfo{address}{Saint-Petersburg, Russia}, \bibinfo{year}{2015}), pp.
  \bibinfo{pages}{126--127}.

\bibitem[{\citenamefont{Gantmakher and Levinson}(2012)}]{gantmakher2012carrier}
\bibinfo{author}{\bibfnamefont{V.}~\bibnamefont{Gantmakher}} \bibnamefont{and}
  \bibinfo{author}{\bibfnamefont{Y.}~\bibnamefont{Levinson}},
  \emph{\bibinfo{title}{Carrier Scattering in Metals and Semiconductors}},
  Modern Problems in Condensed Matter Sciences (\bibinfo{publisher}{Elsevier
  Science}, \bibinfo{year}{2012}).

\bibitem[{\citenamefont{Tarasenko}(2008)}]{tarasenko2008}
\bibinfo{author}{\bibfnamefont{S.~A.} \bibnamefont{Tarasenko}},
  \bibinfo{journal}{Phys. Rev. B} \textbf{\bibinfo{volume}{77}},
  \bibinfo{pages}{085328} (\bibinfo{year}{2008}).

\bibitem[{\citenamefont{Averkiev et~al.}(2002)\citenamefont{Averkiev, Golub,
  and Willander}}]{averkiev2002}
\bibinfo{author}{\bibfnamefont{N.~S.} \bibnamefont{Averkiev}},
  \bibinfo{author}{\bibfnamefont{L.~E.} \bibnamefont{Golub}}, \bibnamefont{and}
  \bibinfo{author}{\bibfnamefont{M.}~\bibnamefont{Willander}},
  \bibinfo{journal}{J. Phys.: Condens. Matter} \textbf{\bibinfo{volume}{14}},
  \bibinfo{pages}{R271} (\bibinfo{year}{2002}).

\bibitem[{\citenamefont{Tarasenko and Ivchenko}(2005)}]{tarasenko2005}
\bibinfo{author}{\bibfnamefont{S.~A.} \bibnamefont{Tarasenko}}
  \bibnamefont{and} \bibinfo{author}{\bibfnamefont{E.~L.}
  \bibnamefont{Ivchenko}}, \bibinfo{journal}{JETP Lett.}
  \textbf{\bibinfo{volume}{81}}, \bibinfo{pages}{231} (\bibinfo{year}{2005}).

\bibitem[{\citenamefont{Ivchenko and Tarasenko}(2004)}]{ivchenko2004}
\bibinfo{author}{\bibfnamefont{E.~L.} \bibnamefont{Ivchenko}} \bibnamefont{and}
  \bibinfo{author}{\bibfnamefont{S.~A.} \bibnamefont{Tarasenko}},
  \bibinfo{journal}{JETP} \textbf{\bibinfo{volume}{99}}, \bibinfo{pages}{379}
  (\bibinfo{year}{2004}).

\bibitem[{\citenamefont{Zholudev et~al.}(2012)\citenamefont{Zholudev, Teppe,
  Orlita, Consejo, Torres, Dyakonova, Czapkiewicz, Wr\'obel, Grabecki,
  Mikhailov et~al.}}]{Zholudev2012}
\bibinfo{author}{\bibfnamefont{M.}~\bibnamefont{Zholudev}},
  \bibinfo{author}{\bibfnamefont{F.}~\bibnamefont{Teppe}},
  \bibinfo{author}{\bibfnamefont{M.}~\bibnamefont{Orlita}},
  \bibinfo{author}{\bibfnamefont{C.}~\bibnamefont{Consejo}},
  \bibinfo{author}{\bibfnamefont{J.}~\bibnamefont{Torres}},
  \bibinfo{author}{\bibfnamefont{N.}~\bibnamefont{Dyakonova}},
  \bibinfo{author}{\bibfnamefont{M.}~\bibnamefont{Czapkiewicz}},
  \bibinfo{author}{\bibfnamefont{J.}~\bibnamefont{Wr\'obel}},
  \bibinfo{author}{\bibfnamefont{G.}~\bibnamefont{Grabecki}},
  \bibinfo{author}{\bibfnamefont{N.}~\bibnamefont{Mikhailov}},
  \bibinfo{author}{\bibfnamefont{S.}~\bibnamefont{Dvoretskii}},
  \bibinfo{author}{\bibfnamefont{A.}~\bibnamefont{Ikonnikov}},
  \bibinfo{author}{\bibfnamefont{K.}~\bibnamefont{Spirin}},
  \bibinfo{author}{\bibfnamefont{V.}~\bibnamefont{Aleshkin}},
  \bibinfo{author}{\bibfnamefont{V.}~\bibnamefont{Gavrilenko}},
  \bibnamefont{and} \bibinfo{author}{\bibfnamefont{W.}~\bibnamefont{Knap}},
  \bibinfo{journal}{Phys. Rev. B} \textbf{\bibinfo{volume}{86}},
  \bibinfo{pages}{205420} (\bibinfo{year}{2012}).

\bibitem[{\citenamefont{Novik et~al.}(2005)\citenamefont{Novik,
  Pfeuffer-Jeschke, Jungwirth, Latussek, Becker, Landwehr, Buhmann, and
  Molenkamp}}]{Novik2005}
\bibinfo{author}{\bibfnamefont{E.~G.} \bibnamefont{Novik}},
  \bibinfo{author}{\bibfnamefont{A.}~\bibnamefont{Pfeuffer-Jeschke}},
  \bibinfo{author}{\bibfnamefont{T.}~\bibnamefont{Jungwirth}},
  \bibinfo{author}{\bibfnamefont{V.}~\bibnamefont{Latussek}},
  \bibinfo{author}{\bibfnamefont{C.~R.} \bibnamefont{Becker}},
  \bibinfo{author}{\bibfnamefont{G.}~\bibnamefont{Landwehr}},
  \bibinfo{author}{\bibfnamefont{H.}~\bibnamefont{Buhmann}}, \bibnamefont{and}
  \bibinfo{author}{\bibfnamefont{L.~W.} \bibnamefont{Molenkamp}},
  \bibinfo{journal}{Phys. Rev. B} \textbf{\bibinfo{volume}{72}},
  \bibinfo{pages}{035321} (\bibinfo{year}{2005}).

\bibitem[{\citenamefont{Sidorov et~al.}(2001)\citenamefont{Sidorov, Dvoretskii,
  Varavin, Mikhailov, Yakushev, and Sabinina}}]{Sidorov2001}
\bibinfo{author}{\bibfnamefont{Y.~G.} \bibnamefont{Sidorov}},
  \bibinfo{author}{\bibfnamefont{S.~A.} \bibnamefont{Dvoretskii}},
  \bibinfo{author}{\bibfnamefont{V.~S.} \bibnamefont{Varavin}},
  \bibinfo{author}{\bibfnamefont{N.~N.} \bibnamefont{Mikhailov}},
  \bibinfo{author}{\bibfnamefont{M.~V.} \bibnamefont{Yakushev}},
  \bibnamefont{and} \bibinfo{author}{\bibfnamefont{I.~V.}
  \bibnamefont{Sabinina}}, \bibinfo{journal}{Semiconductors}
  \textbf{\bibinfo{volume}{35}}, \bibinfo{pages}{1045} (\bibinfo{year}{2001}).

\bibitem[{\citenamefont{Varavin et~al.}(1995)\citenamefont{Varavin, Dvoretsky,
  Liberman, Mikhailov, and Sidorov}}]{Varavin1995}
\bibinfo{author}{\bibfnamefont{V.}~\bibnamefont{Varavin}},
  \bibinfo{author}{\bibfnamefont{S.}~\bibnamefont{Dvoretsky}},
  \bibinfo{author}{\bibfnamefont{V.}~\bibnamefont{Liberman}},
  \bibinfo{author}{\bibfnamefont{N.}~\bibnamefont{Mikhailov}},
  \bibnamefont{and} \bibinfo{author}{\bibfnamefont{Y.}~\bibnamefont{Sidorov}},
  \bibinfo{journal}{Thin Solid Films} \textbf{\bibinfo{volume}{267}},
  \bibinfo{pages}{121} (\bibinfo{year}{1995}).

\bibitem[{\citenamefont{Bel'kov et~al.}(2005)\citenamefont{Bel'kov, Ganichev,
  Ivchenko, Tarasenko, Weber, Giglberger, Olteanu, Tranitz, Danilov, Schneider
  et~al.}}]{Belkov2005}
\bibinfo{author}{\bibfnamefont{V.~V.} \bibnamefont{Bel'kov}},
  \bibinfo{author}{\bibfnamefont{S.~D.} \bibnamefont{Ganichev}},
  \bibinfo{author}{\bibfnamefont{E.~L.} \bibnamefont{Ivchenko}},
  \bibinfo{author}{\bibfnamefont{S.~A.} \bibnamefont{Tarasenko}},
  \bibinfo{author}{\bibfnamefont{W.}~\bibnamefont{Weber}},
  \bibinfo{author}{\bibfnamefont{S.}~\bibnamefont{Giglberger}},
  \bibinfo{author}{\bibfnamefont{M.}~\bibnamefont{Olteanu}},
  \bibinfo{author}{\bibfnamefont{H.-P.} \bibnamefont{Tranitz}},
  \bibinfo{author}{\bibfnamefont{S.~N.} \bibnamefont{Danilov}},
  \bibinfo{author}{\bibfnamefont{P.}~\bibnamefont{Schneider}},
  \bibinfo{author}{\bibfnamefont{W.}~\bibnamefont{Wegscheider}},
  \bibinfo{author}{\bibfnamefont{D.}~\bibnamefont{Weiss}}, \bibnamefont{and}
  \bibinfo{author}{\bibfnamefont{W.}~\bibnamefont{Prettl}},
  \bibinfo{journal}{J. Phys.: Condens. Matter} \textbf{\bibinfo{volume}{17}},
  \bibinfo{pages}{3405} (\bibinfo{year}{2005}).

\bibitem[{\citenamefont{Ivchenko}(2005)}]{Ivchenkobook}
\bibinfo{author}{\bibfnamefont{E.~L.} \bibnamefont{Ivchenko}},
  \emph{\bibinfo{title}{Optical Spectroscopy of Semiconductor Nanostructures}}
  (\bibinfo{publisher}{Alpha Science International Ltd}, \bibinfo{year}{2005}).

\bibitem[{\citenamefont{Ivchenko and Ganichev}(2017)}]{IvchenkoGanichev2017}
\bibinfo{author}{\bibfnamefont{E.}~\bibnamefont{Ivchenko}} \bibnamefont{and}
  \bibinfo{author}{\bibfnamefont{S.}~\bibnamefont{Ganichev}},
  \emph{\bibinfo{title}{Spin Photogalvanics}} (\bibinfo{publisher}{Springer ed.
  M.I. Dyakonov}, \bibinfo{year}{2017}).

\bibitem[{\citenamefont{Ganichev et~al.}(2001)\citenamefont{Ganichev, Ivchenko,
  Danilov, Eroms, Wegscheider, Weiss, and Prettl}}]{Ganichev2001}
\bibinfo{author}{\bibfnamefont{S.~D.} \bibnamefont{Ganichev}},
  \bibinfo{author}{\bibfnamefont{E.~L.} \bibnamefont{Ivchenko}},
  \bibinfo{author}{\bibfnamefont{S.~N.} \bibnamefont{Danilov}},
  \bibinfo{author}{\bibfnamefont{J.}~\bibnamefont{Eroms}},
  \bibinfo{author}{\bibfnamefont{W.}~\bibnamefont{Wegscheider}},
  \bibinfo{author}{\bibfnamefont{D.}~\bibnamefont{Weiss}}, \bibnamefont{and}
  \bibinfo{author}{\bibfnamefont{W.}~\bibnamefont{Prettl}},
  \bibinfo{journal}{Phys. Rev. Lett.} \textbf{\bibinfo{volume}{86}},
  \bibinfo{pages}{4358} (\bibinfo{year}{2001}).

\end{thebibliography}

\end{document}